\documentclass[review]{elsarticle}

\usepackage{lineno,hyperref}
\modulolinenumbers[5]

\usepackage{dirtytalk}
\usepackage{amsmath}
\usepackage{amsfonts}
\usepackage{subcaption}
\usepackage{mathabx}
\usepackage{xcolor}

\usepackage{algorithm}
\usepackage[noend]{algpseudocode}
\journal{Journal of \LaTeX\ Templates}

\biboptions{sort&compress}
\usepackage{nomencl}
\makenomenclature

\begin{document}

\begin{frontmatter}

\title{Reduced basis methods for numerical room acoustic simulations with parametrized boundaries}

\author[mymainaddress,mymainaddress2]{Hermes Sampedro Llopis}
\ead{hsllo@elektro.dtu.dk}

\author[mymainaddress3]{Allan P. Engsig-Karup}
\author[mymainaddress2]{Cheol-Ho Jeong}
\author[mymainaddress4]{Finnur Pind}
\author[mymainaddress5]{Jan S. Hesthaven}

\address[mymainaddress]{Rambøll Denmark, Copenhagen, Denmark}
\address[mymainaddress2]{Acoustic Technology Group, Department of Electrical Engineering, Technical University of Denmark,
Kongens Lyngby, Denmark}
\address[mymainaddress3]{Scientific Computing Section, Department of Applied Mathematics and Computer Science, Technical University of Denmark, Kongens Lyngby, Denmark}
\address[mymainaddress4]{Treble Technologies, Reykjavík, Iceland}
\address[mymainaddress5]{Chair of Computational Mathematics and Simulation Science, Ecole Polytechnique Federale de Lausanne,
Lausanne, Switzerland}

\begin{abstract}
The use of model-based numerical simulation of wave propagation in rooms for engineering applications requires that acoustic conditions for multiple parameters are evaluated iteratively and this is computationally expensive. We present a reduced basis methods (RBM) to achieve a computational cost reduction relative to a traditional full order model (FOM), for wave-based room acoustic simulations with parametrized boundary conditions. In this study, the FOM solver is based on the spectral element method, however other numerical methods could be applied. The RBM reduces the computational burden by solving the problem in a low-dimensional subspace for parametrized frequency-independent and frequency-dependent boundary conditions. The problem is formulated and solved in the Laplace domain, which ensures the stability of the reduced order model based on the RBM approach. We study the potential of the proposed RBM framework in terms of computational efficiency, accuracy and storage requirements and we show that the RBM leads to 100-fold speed-ups for a 2D case with an upper frequency of 2kHz and around 1000-fold speed-ups for an analogous 3D case with an upper frequency of 1kHz. While the FOM simulations needed to construct the ROM are expensive, we demonstrate that despite this cost, the ROM has a potential of three orders of magnitude faster than the FOM when four different boundary conditions are simulated per room surface. Moreover, results show that the storage model for the ROM is relatively high but affordable for the presented 2D and 3D cases.
\end{abstract}

\begin{keyword}
Room acoustic simulations; model order reduction; reduced basis methods; Laplace domain; complex boundaries.
\end{keyword}

\end{frontmatter}

\linenumbers

\section{\label{sec:1} Introduction}
Room acoustic simulations are used for various purposes, e.g., building design, music, hearing research, entertainment, and virtual reality (VR). Historically, these simulations have mostly been carried out by means of geometrical acoustics (GA) methods \cite{SaviojaSvensson,Vorlaender1}, such as ray tracing, which approximate the sound propagation to ensure a manageable computational cost, but fails to simulate the correct wave nature of sound, resulting in diffraction and interference at low frequencies. The geometric approximation is known to cause a considerable degradation of the simulation accuracy in many cases \cite{georoundrobin}. Another approach is to numerically solve the governing equations, i.e., the wave equation in the time-domain or the Helmholtz equation in the frequency domain, using numerical discretization methods. Different numerical methods have been applied to the room acoustics problem in the past, e.g., the finite element method (FEM) \cite{FEM,Okuzono1}, the spectral element method (SEM) \cite{SEM}, the finite difference method (FDTD) \cite{FDTS}, the boundary element method (BEM) \cite{BEM} and the discontinuous Galerkin finite element method (DG-FEM) \cite{dgfemWang}. These numerical methods are, in principle, more accurate than GA, as no approximation on the wave propagation is introduced except for numerical errors by the discretization. 

The main drawback of a wave-based modelling approach is the high computational cost, especially when modeling large spaces at higher frequencies. This makes it difficult to apply them to scenarios where the full audible spectrum and large room dimensions must be considered, e.g., simulating a large hall from 20 Hz to 20 kHz. A wave-based simulation of a large hall up to 20 kHz remains very challenging, and the goal of this study is not to propose a methodology that enables this. Instead, we consider the case where the same room is simulated multiple times under varying boundary conditions. This type of use case is very common in e.g. building design, where different surface materials are tested to identify desired acoustic conditions. Thus, reducing the computation time for the parameterized boundary condition case is of utmost importance for many applications.

We propose a computational framework to reduce the computational cost compared to traditional \textit{full-order} numerical solutions. The framework is based on a reduced basis method (RBM) for the acoustics problem that includes parametrization of key design parameters at the boundary of the domain. It can be applied e.g. to surface absorption and scattering, and enables a substantial cost reduction when simulating a room multiple times under parameter variation at the boundaries.

Any full-order numerical solver can be used in the RBM framework to capture the underlying physics needed to build the reduced basis. It is relevant to impose the following requirements; 1) geometric flexibility, which allows simulation of curved and complex geometries, 2) high-order accuracy to minimize wave dispersion over long distances, 3) transient response reconstruction, i.e., impulse response, 4) a stable reduced order model (ROM) and 5) the possibility to extend the applicability to large spaces and high frequencies. To fulfill 1) methods such as FDTD with structured grids are not well suited. Instead, FEM, BEM, SEM or DG-FEM provide geometric flexibility, which is needed in room acoustics to accurately model complex shaped rooms, i.e. curved geometries \cite{SEM}. Requirement 2) can be fulfilled by any high-order scheme. SEM and DG-FEM are examples of numerical methods that lend themselves naturally to high-order discretizations \cite{KreissOliger} with low dispersion and dissipation properties \cite{Dips_SEM2,Dips_SEM3,Dips_SEM1}.
To fulfill 3) time-domain (TD) and Laplace domain (LD) are more suitable than Fourier-domain (FD). In particular, when solving the problem in the frequency domain, LD is more appropriate for non-periodic problems with a transient response. In terms of requirement 4) TD approaches are known to be challenging when it comes to stability robustness of reduced order modeling frameworks \cite{stab1,stab2,stab3,stab4}. Finally, 5) requires scalable methods, ideally of $\mathcal{O}(N)$, where $N$ is the total degrees of freedom in the scheme. Considering these desired features, we have chosen to work with the spectral element method for the full-order model solver in the Laplace domain. When compared against DG-FEM, the SEM is simpler when solving in the Laplace domain, while DG-FEM requires definition of the fluxes \cite{cockburn1,cockburn2} and results in larger discrete problems. But we highlight again that any numerical framework could be used for the full order model.


RBM represents an emerging field of numerical techniques for efficiently solving parametrized problems when a large number of model-based simulations with different parameter values are needed \cite{CRBM,CRBM2, NgocCRBM,RozzaHuynhRBM,DrohmannRBM, PrudRBM, QuarteroniRBM,HolmesTurb,RBMSuccessLimitations}. RBM have been successfully applied in many different fields, e.g., computational fluid dynamics \cite{AmsallemFluid, phdGiere}, electromagnetics \cite{Ganeshele, Chenelec}, heat transfer \cite{phdGrepl} and vibroacoustics \cite{SrinivasaVA, HetmaniukVA, HerrmannVA}. However, the method has been applied to the wave equation only in a few studies \cite{MMiller, Bigoni, AfkhamWE, PereyraWE,PereyraWE2,PereyraWE3}. The RBM consists of two stages; An offline stage where the parameter space is explored to generate a problem-dependent basis using the full order model solver, and a Galerkin projection to reduce the dimensionality of the problem by utilizing the generated basis. The Galerkin projection requires access to the code, hence, it is referred to as an intrusive technique. An online stage, enables the problem to be solved for the new parameter value at a much lower computational cost by evaluating the reduced problem. The generation of the basis can be seen as a data-driven technique that relies on a proper orthogonal decomposition (POD). This offline stage is typically computationally costly as it requires multiple full order solutions, to produce so-called ``snapshots'' that capture relevant states for different parameters. The reduction of the size of the computational problem then comes with the truncation of the basis, which may be a source of numerical instability in the reduced model even when the full order model (FOM) is stable \cite{stab1}. A critical challenge in RBM techniques is the instability issue that may appear in long time numerical simulations if not taken care of in the formulation of the method. Some remedies have been proposed to address the instability in the time-domain \cite{Moore, Amsallem, Kalashnikova,AfkhamWE}. A previous study solves and reduces the acoustic-elastic wave equation in the Laplace domain to ensure stability \cite{Bigoni}. One challenge for the Laplace domain method is the time signal reconstruction, which is important in acoustics for computing the different acoustic parameters \cite{ISO3382}, playback and auralization purposes \cite{Vorlaender1}. In Addition, the combination of RBM and SEM, has been investigated in other areas, e.g., for the Navier-Stokes equations \cite{rozza1,rozza2}, where a potential synergy between high-order discretization and reduced basis methods is demonstrated.

The main novelty and contribution of this study is twofold:
1) it derives a 2D and 3D RBM framework formulated in the Laplace domain for the room acoustic problem with parameterized frequency-independent and frequency-dependent boundary conditions and  2) it provides a detailed analysis of the potential and drawbacks of the framework for room acoustic simulations.




\section{Method description}
\subsection{Full Order Model in Laplace domain}
The acoustic wave propagation in a lossless and steady medium is described by the second-order wave equation,
\begin{equation}\label{waveeq}
    \frac{\partial^2 \hat{p}}{\partial t^2} - c^2\Delta  \hat{p} = 0,\quad \textrm{in } \Omega \times (0,T],
\end{equation}
where $\hat{p}(\boldsymbol{x},t)$ is the sound pressure, $\boldsymbol{x}\in \Omega$ the position in the domain $\Omega\subset 	\mathbb{R}^d$ with $d={2,3}$, $t$ is the time in the interval $(0, T]$ s and $c$ is the speed of sound ($c = 343$ m/s). The second-order equation can also be written as a system of two coupled linear first-order partial differential equations
\begin{subequations}
\begin{align}
     \frac{\partial \boldsymbol{\hat{v}}}{\partial t}&=-\frac{1}{\rho}\nabla \hat{p}, \qquad \textrm{in } \Omega \times (0,T],\\
    \frac{\partial \hat{p}}{\partial t} &=-\rho c^2\nabla \cdot \boldsymbol{\hat{v}}, 
\end{align}
\end{subequations}
where $\boldsymbol{\hat{v}}(\boldsymbol{x},t)$ is the particle velocity and $\rho$ is the density of the medium ($\rho=1.2$ kg/m$^3$). The system can be excited with an initial condition, e.g., a Gaussian pulse with a spatial variance $\sigma_g$ for the sound pressure, and zero for the particle velocity.

In this study, a Laplace domain approach is considered. The time-dependent problem can be reduced to the computation of the Laplace transform of the sound pressure evaluated at a fixed complex frequency $s=\sigma + i\gamma$, by multiplying (\ref{waveeq}) by $e^{-st}$ and integrating in time over the interval $[0,\infty)$ to obtain
\begin{equation}\label{welap}
    s^2p- c^2 \Delta p = s\hat{p}_0+\hat{p}_{t_{,0}}, 
\end{equation}
where $p(\boldsymbol{x},s)$ is the Laplace transform of $\hat{p}(\boldsymbol{x},t)$, $\hat{p}_0(\boldsymbol{x)}=\hat{p}(\boldsymbol{x},t=0)$ is the initial condition in the time-domain and $\hat{p}_{t_{,0}}(\boldsymbol{x)}=\left.\frac{\partial \hat{p}(\boldsymbol{x},t)}{\partial t}\right|_{t=0}$. A Gaussian pulse is considered,
\begin{equation}\label{initial_cond}
    \hat{p}_0(\boldsymbol{x}) = e^{-\Big(\frac{(\boldsymbol{x}-\boldsymbol{x}_0)^2}{\sigma_g^2} \Big)}, \quad \hat{p}_{t_{,0}}(\boldsymbol{x}) = 0,
\end{equation}
where $\boldsymbol{x}_0$ is the position of the source and $\sigma_g$ is the spatial variance.  Note that the Laplace transform of $\hat{\boldsymbol{v}}(\boldsymbol{x},t)$ will be denoted by ${\boldsymbol{v}}(\boldsymbol{x},s)$. 

To solve (\ref{welap}) numerically, the spatial derivatives are discretized and each complex frequency is evaluated separately.
By multiplying (\ref{welap}) with a test function $w$, integrating over the domain and making use of Green's first identity, we obtain the weak formulation

\begin{align}\label{weekfG}
    \int_{\Omega}s^2pw d\Omega   &+\int_{\Omega}  c^2\nabla  p\cdot\nabla w d\Omega \\ \nonumber
    &-\oint_{\Gamma}c^2\frac{\partial p}{\partial \mathbf{n}}w d\Gamma =\int_{\Omega}s\hat{p}_0wd\Omega,
\end{align}
where $\mathbf{n}$ is the outward pointing normal vector of the boundary $\Gamma$ of $\Omega$. For harmonic signals, the conservation of momentum yields a proportional relation between the surface normal pressure derivative ($ \frac{\partial p}{\partial \mathbf{n}}$)  and the surface normal velocity at the boundary ($ v_n=v \cdot\mathbf{n}$) given by $ \frac{\partial p}{\partial \mathbf{n}} = -s\rho v_n$. In room acoustics it is common to define the boundary conditions in terms of the surface impedance $Z_s$ \cite{Kuttruff}. The normal incidence absorption coefficient $\alpha_{norm}$ can be obtained from the surface impedance as follows
\begin{equation}\label{abscoeff}
    \alpha_{norm} = 1-\Bigg|\frac{Z_s-\rho c}{Z_s+\rho c}\Bigg|^2.
\end{equation}
Given the relation $Z_s=p_{\Gamma}/v_n$, the impedance boundary condition can be written
\begin{equation}\label{impBC}
    \frac{\partial p}{\partial \mathbf{n}} = -s\rho\frac{p_{\Gamma}}{Z_s},
\end{equation}
where $p_{\Gamma}$ denotes the sound pressure at the boundary. Substituting (\ref{impBC}) into (\ref{weekfG}) allows elimination of the normal pressure in the boundary integrals
\begin{align}\label{weekf12}
     \int_{\Omega}s^2pw d\Omega   &+\int_{\Omega}  c^2\nabla p\cdot\nabla w d\Omega \\ \nonumber
     &+ \oint_{\Gamma}sc^2\frac{\rho}{Z_s}p_{\Gamma} wd\Gamma = \int_{\Omega}s\hat{p}_0wd\Omega.
\end{align}
The SEM formulation is well known, an overview and the definition of the operators can be found  \cite{NODALDGFEM,SEM,SPHPAllan}. The formulation written in the Laplace domain is given by
\begin{align}\label{wedisc}
 \big(s^2\mathcal{M}+c^2\mathcal{S} + sc^2\frac{\rho}{Z_s}\mathcal{M}_{\Gamma}\big)\mathbf{p} =s\hat{p}_0\mathcal{M},
\end{align}
where $\mathcal{M}\in \mathbb{R}^{N \times N}$ is the mass matrix, $\mathcal{S}\in \mathbb{R}^{N \times N}$ is the stiffness matrix and $N$ denotes the degrees of freedom (DOF). The stiffness matrix for 2D is given by $\mathcal{S}=\mathcal{S}_x + \mathcal{S}_y$ and for 3D is $\mathcal{S}=\mathcal{S}_x+\mathcal{S}_y+\mathcal{S}_z$. Here the $x,y,z$ subscripts denote differentiation in Cartesian directions, respectively.
It is important for the implementation to split (\ref{wedisc}) into a set of $2N$ real equations for a given frequency $s = \sigma + i\gamma\in \mathbb{C}$, where the solution is written  as $\boldsymbol{p} = \boldsymbol{p}^{\sigma} + i\boldsymbol{p}^{\gamma}$ \cite{Bigoni}. After simple manipulations, the system can be written as
\begin{align}\label{eqinarow}
     &(\mathbf{K}^{\sigma}\boldsymbol{p}^{\sigma}-\mathbf{K}^{\gamma}\boldsymbol{p}^y)+i(\mathbf{K}^{\gamma}\boldsymbol{p}^{\sigma}+\mathbf{K}^{\sigma}\boldsymbol{p}^{\gamma})=\boldsymbol{ Q}^{\sigma} + i\boldsymbol{Q}^{\gamma},
\end{align}
where
\begin{subequations}
\begin{align}
    &\mathbf{K}^{\sigma} = (\sigma^2 - \gamma^2)\mathcal{M} +c^2\mathcal{S} + \sigma\mathcal{B},\\ &\mathbf{K}^{\gamma} = 2\sigma \gamma\mathcal{M}+\gamma\mathcal{B},\\
   &\boldsymbol{ Q}^{\sigma} = \sigma \hat{p}_0\mathcal{M}, \\ &\boldsymbol{Q}^{\gamma} = \gamma \hat{p}_0\mathcal{M}\\
   &\mathcal{B} = c^2\frac{\rho}{Z_s}\mathcal{M}_{\Gamma}.
\end{align}
\end{subequations}
Finally, (\ref{eqinarow}) can be written as
\begin{gather}\label{Generalmatrix}
 \begin{bmatrix} \mathbf{K}^{\sigma} & -\mathbf{K}^{\gamma} \\ \mathbf{K}^{\gamma} & \mathbf{K}^{\sigma} \end{bmatrix}
  \begin{bmatrix} \boldsymbol{p}^{\sigma} \\ \boldsymbol{p}^{\gamma}\end{bmatrix}
 = \begin{bmatrix} \boldsymbol{Q}^{\sigma} \\ \boldsymbol{Q}^{\gamma}\end{bmatrix}.
\end{gather} 
Note that for perfectly rigid boundaries, the particle velocity at the boundaries is $v_n=0$, thus, $\mathcal{B}=0$ and (\ref{wedisc}) reduces to
\begin{align}\label{hardwalls}
 \big(s^2\mathcal{M}+c^2\mathcal{S}\big)\mathbf{p} = s\hat{p}_0\mathcal{M}.
\end{align}


\subsubsection{Frequency-dependent boundary condition}
The frequency-dependent boundary conditions are implemented via the method of auxiliary differential equations (ADE) \cite{Cotte,Troian,SEM}. The sound pressure $p_{\Gamma}(\omega)$ and the normal velocity $v_n(\omega)$ at the boundaries are related through the relationship in the Fourier-domain
\begin{equation}
    v_n(\omega) = \frac{p_{\Gamma}(\omega)}{Z_s(\omega)} = p_{\Gamma}(\omega)Y_s(\omega),
\end{equation}
where $\omega$ is the angular frequency and $Z_s(\omega)$ is the surface impedance at the boundary. The admittance at the boundary $Y_s(\omega)$, is the inverse of $Z_s(\omega)$, which can be approximated as a rational function of order $Q$ as

\begin{equation}\label{rationalfun}
    Y_s(\omega)=\frac{a_0+...+a_Q(-i\omega)^Q}{1+...+b_Q(-i\omega)^Q},
\end{equation}
and can be written as

\begin{align}\label{Ysaprox}
    Y_s(\omega)&= Y_{\infty} +\sum_{k=1}^L\frac{A_k}{\hat{\lambda}_k - i\omega} \\ \nonumber
    &+ \sum_{k=1}^S \left(\frac{B_k+iC_k}{\hat{\alpha}_k+i\hat{\beta}_k-i\omega} + \frac{B_k-iC_k}{\hat{\alpha}_k-i\hat{\beta}_k-i\omega}\right).
\end{align}
Hence $L$ is the number of real poles $\hat{\lambda}_k$, $S$ are the number of complex conjugate pole pairs  $\hat{\alpha}_k\pm i\hat{\beta}_k$, used in the rational function approximation, and $Y_{\infty}$, $A_k$, $B_k$ and $C_k$ are coefficients. Applying the inverse Fourier transform and inserting the particle velocity definition, we recover the final expression
\begin{align}\label{verloc}
    \hat{v}_n(t) = Y_{\infty}\hat{p}_{\Gamma}(t) &+ \sum_{k=1}^L A_k\hat{\phi}_k(t) \\ \nonumber
    &+ \sum_{k=1}^S 2\Big[B_k\hat{\psi}_k^{(1)}(t)+C_k\hat{\psi}_k^{(2)}(t)\Big],
\end{align}
where $\hat{\phi}_k$, $\hat{\psi}_k^{(1)}$ and $\hat{\psi}_k^{(2)}$ are the so called accumulators that can be determined by solving the corresponding differential equations \cite{Troian}. Transforming the expressions to the Laplace domain yields
\begin{subequations}
\begin{align}
     & \hat{\phi}_k(s)=p_{\Gamma}(s)(s + \hat{\lambda}_k)^{-1},\\
     & \hat{\psi}_k^{(2)}(s) = \hat{\beta}_k p_{\Gamma}(s) \big[(s+\hat{\alpha}_k)^2+\hat{\beta}_k^2\big]^{-1} ,\\ \label{psi1acum}
     &\hat{\psi}_k^{(1)}(s)=(s+\hat{\alpha}_k)\hat{\beta}_k^{-1}\hat{\psi}_k^{(2)}(s).
\end{align}
\end{subequations}
The auxiliary differential equations to be solved can be written in matrix form for the real and imaginary parts by considering $s=\sigma+i\gamma$

\begin{gather}\label{matrix1}
 \begin{bmatrix} \mathbf{G}^{\sigma} & -\mathbf{G}^{\gamma} \\ \mathbf{G}^{\gamma} & \mathbf{G}^{\sigma} \end{bmatrix}
  \begin{bmatrix} \hat{\phi}^{\sigma} \\  \hat{\phi}^{\gamma}\end{bmatrix}
 = \begin{bmatrix} p_{\Gamma}^{\sigma} \\ p_{\Gamma}^{\gamma}\end{bmatrix},
\end{gather}  
where $\mathbf{G}^{\sigma} = \sigma + \hat{\lambda}_k$ and $\mathbf{G}^{\gamma}  = \gamma$. Moreover, 
\begin{gather}\label{matrix2}
 \begin{bmatrix} \mathbf{G}^{\sigma} & -\mathbf{G}^{\gamma} \\ \mathbf{G}^{\gamma} & \mathbf{G}^{\sigma} \end{bmatrix}
  \begin{bmatrix} \hat{\psi}^{(2)\sigma} \\  \hat{\psi}^{(2)\gamma}\end{bmatrix}
 =\hat{\beta} \begin{bmatrix} p_{\Gamma}^{\sigma} \\ p_{\Gamma}^{\gamma}\end{bmatrix},
\end{gather}  
where $\mathbf{G}^{\sigma} = \sigma^2-\gamma^2 +2\alpha\sigma + \hat{\alpha}^2+\hat{\beta}^2$ and $\mathbf{G}^{\gamma} = 2\sigma \gamma+2\hat{\alpha} \gamma.$ Finally, $\hat{\psi}^{(1)}(s)$ can be obtained by substituting $\hat{\psi}^{(2)}(s)$ into (\ref{psi1acum}),
\begin{subequations}\label{psieq}
\begin{align}
    &\hat{\psi}^{(1)\sigma} = \frac{1}{\hat{\beta}}\Big[(\sigma+\hat{\alpha})\hat{\psi}^{(2)\sigma} - \gamma\hat{\psi}^{(2)\gamma}\Big],\\
     &\hat{\psi}^{(1)\gamma} = \frac{1}{\hat{\beta}}\Big[\gamma\hat{\psi}^{(2)\sigma}+(\sigma+\hat{\alpha})\hat{\psi}^{(2)\gamma}\Big].
\end{align}
\end{subequations}
Note that there is not a direct way to transform from the Fourier-domain to the Laplace domain. Thus, a procedure to solve the system, including the frequency-dependent boundaries, is described in Algorithm \ref{alg1}. First, the pressure is obtained from (\ref{hardwalls}). Second, the real and imaginary part of the particle velocity at the boundaries is calculated by (\ref{verloc}). Third, the admittance at the boundaries  is  calculated using the computed pressure and normal velocity at the boundary
\begin{align}\label{BCImpedaceReal}
    Y_s &= Y_s^{\sigma}+iY_s^{\gamma}=\frac{v_{n}}{p_{\Gamma}} \\ \nonumber
    &= \frac{v_{n}
^{\sigma}p_\Gamma^{\sigma}+v_{n}
^{\gamma}p_\Gamma^{\gamma}}{(p^{{\sigma}}_{\Gamma})^2+(p^{\gamma}_{\Gamma})^2} + i \frac{v_{n}
^{\gamma} p_\Gamma^{\sigma}-v_{n\Gamma}
^{\sigma}p_\Gamma^{\gamma}}{(p^{{\sigma}}_{\Gamma})^2+(p^{\gamma}_{\Gamma})^2}.
\end{align}
 In this study we consider porous materials as frequency-dependent boundary conditions. The surface impedance can be estimated using Miki's model \cite{Mikis} in conjunction with a transfer matrix method \cite{allardbook}, and mapped to a four pole rational function by using a vector fitting algorithm \cite{vecfitt}. Finally, the sound pressure in the domain is solved with (\ref{Generalmatrix}) where
\begin{subequations}\label{freqdepall}
\begin{align}\label{freqdep1}
    &\mathbf{K}^{\sigma} = (\sigma^2 - \gamma^2)\mathcal{M} +c^2\mathcal{S}+\sigma\mathcal{B}^{\sigma},\\
   &\mathbf{K}^{\gamma} = 2\sigma y\mathcal{M}+\gamma\mathcal{B}^{\gamma},\\
   &\mathcal{B}^{\sigma} = c^2\rho \big(\sigma Y_{s}^{\sigma} - \gamma Y_{s}^{\gamma} \big)\mathcal{M}_{\Gamma},\label{boundaryImag}\\
&\mathcal{B}^{\gamma} = c^2\rho \big( \gamma Y_{s}^{\sigma} + \sigma Y_{s}^\gamma \big)\mathcal{M}_{\Gamma}.
\end{align}
\end{subequations}

\begin{algorithm}
	\caption{Frequency-dependent boundary conditions} 
	\begin{algorithmic}[1]
	\Procedure{Solver}{}
		\State Approximate $Y_s$  (Miki's model) using the \textit{vectfit}3 algorithm (\ref{Ysaprox})
		\For {$n=1$ to $N_s$}
			
				\State Compute $p_{\Gamma}^{\sigma}$ and  $p_{\Gamma}^{\gamma}$, the pressure at the boundary defined in (\ref{hardwalls})
				\State Compute the accumulators by solving (\ref{matrix1}), (\ref{matrix2}) and (\ref{psieq})
				\State Compute $v_{n}^{\sigma}$ and  $v_{n}^{\gamma}$ defined in (\ref{verloc})
				\State Compute the admittance at the boundary $Y_s$ given in (\ref{BCImpedaceReal})
				\State Compute the pressure $p$ by solving (\ref{Generalmatrix})

		\EndFor
		\Return $p^{\sigma}$, $p^{\gamma}$
		\EndProcedure
	\end{algorithmic} \label{alg1}
\end{algorithm}
Note that since $Y_s$ has the same value in the whole boundary, it is sufficient to compute steps 1 to 7 for a single boundary node.


\subsection{Time-domain signal reconstruction}
In room acoustics the impulse responses are widely used to characterize a room for a given source and receiver pair. Temporal representations of audio signals are crucial in room acoustic simulations, particularly for playback and auralization purposes. Moreover, the impulse response is used to extract different room acoustic parameters \cite{ISO3382}. The time reconstruction can be performed with the inverse Laplace transform. Since the analytical solution of the Laplace inverse transform is difficult to evaluate, the Weeks method is used to recover the time-dependent signal \cite{Weeks}. The temporal reconstruction by Weeks method depends on free parameters $(\sigma,b)$ and number of complex evaluated frequencies $N_s$. The time signal can be expressed following the same approach as described in \cite{Bigoni}
\begin{equation}\label{ureco}
    \hat{p}(t) = e^{(\sigma-b)t}\sum_{k=0}^{N_s-1}\hat{a}_kL_k(2bt),
\end{equation}
where $t$ denotes a time instant, $L_k(\cdot)$ is the Laguerre polynomial of degree $k = 0,...,N_s-1$, which can be computed recursively using Clenshaw's algorithm \cite{Clenshaw}. The approximate expansion coefficients denoted by $\hat{a}_k$, depend on the Laplace solution and can be approximated as
\begin{equation}\label{weekscoef}
    \hat{a}_{k} = \frac{b}{N_z}\sum_{j=-N_s}^{N_s-1}\frac{e^{-ik\theta_{j+1/2}}}{1-e^{i\theta_{j+1/2}}} p(s_j),
\end{equation}
where $k = 0,...,N_s-1$.  
The complex frequency is given by
\begin{subequations}
\begin{align}
       &s_j = \sigma +i\gamma= \sigma + ibg(\theta), \\
        &g(\theta)=\frac{\cot(\theta_{j+1/2})}{2},\quad \textrm{for } j = -N_s,...,N_s-1,
\end{align}\label{compelexfreq}
\end{subequations}
where $\theta_j=j\pi/N_s$, $\theta_{j+1/2}=(j+1/2)\pi/N_s$. To improve the efficiency of the solver, two different polynomial orders can be chosen depending on the complex frequency
\begin{align}
   &P_{l} \quad\textrm{for }\quad |g(\theta)|\simeq 0,&\quad&
   P_{h} \quad\textrm{for }\quad |g(\theta)|>\!\!> 1.
\end{align}
By trigonometric identities it can be shown that by computing only the solution for $j=-N_s,...,-1$, a reduction in the number of required solutions is possible. Thus, the expression of $p(s_j)$ in (\ref{weekscoef}) can be substituted  by
\begin{subequations}
\begin{align}\label{wertimerec}
    p(s_j) &= p^{\sigma}_{j} + i p^{\gamma}_{j}, \quad \textrm{for} \quad j=0,...,N_s-1,\\
    p(s_k) &= p^{\sigma}_{k} - i p^{\gamma}_{k}, \quad \textrm{for}\quad k=2N_s-j+1,
\end{align}
\end{subequations}
where $j=0,...,N_s-1$.

 The choice of the free parameters $(\sigma, b)$ can compromise the final results. A suboptimal choice may lead to large deviations from the actual solution. Moreover, choosing larger values of $N_s$ than optimal, would avoid deviations but may lead to a longer computational time.  There are different rules of thumb to find the optimum values for these parameters \cite{Piessens, Garbow}, but as discussed in \cite{Bigoni}, none of them are convenient for this particular case. Following the approach presented in \cite{Bigoni}, the parameters are chosen by fixing a given resolution $N_s$, a number of time steps $N_t$ and computing the solution for different parameter values that are compared with an analytical solution or a high fidelity time solver solution \cite{SEM}. Hence
\begin{equation}\label{optimization}
   \sigma^{opt},b^{opt} = \min_{\sigma,b} \Bigg|\Bigg| \sum_i\big(\hat{p}^*_i(\boldsymbol{x},t)-p^*_i(\boldsymbol{x},t)\big)\Bigg|\Bigg|_2^2,
\end{equation}
   where $\hat{p}^*_i(\boldsymbol{x},t)$ is the solution obtained with a time solver and $p^*_i(\boldsymbol{x},t)$ is the solution obtained by (\ref{ureco}).
A way to determine the number of complex frequencies $N_s$ is to select the value that provides the same error as obtained with the time solver. Once the parameters are found for the FOM, they can be used also for the RBM.

 
   \subsection{Reduced Order Model}
   
The main motivation for reduced basis methods is to obtain a sufficiently accurate solution to a parametrized problem, for any value of the given parameter, at a reduced computational cost compared to the original high-fidelity solver. The method consists of two stages. First an offline stage a set of high-fidelity solutions, called snapshots, are computed for a selected set of parameter values to capture the relevant physics. The evaluation of the FOM for multiple parameter values results in a upfront computational cost and is used to generate $N_{rb}$ basis functions by using, e.g., Proper Orthogonal Decomposition (POD). A Galerkin projection of the original problem results in a reduced problem with $N_{rb}\ll N$ degrees of freedom. During the online stage a system of dimension $N_{rb}$ is solved for each new parameter value and the solution is recovered by a linear combination of the pre-generated basis functions. The key challenge of the method is to construct a reduced basis that preserves the physics dynamics for a required accuracy level.
The high-fidelity solutions $p_{fb}(\boldsymbol{x},s,\mu)=p^T$ of the parametrized problem under the variation of the parameter $\mu$, belonging to the parameter space $\mathbb{P}$, approximates the solution manifold
\begin{equation}
    \textit{M} =\{p(\mu)|\mu \in \mathbb{P} \},
\end{equation}
The ultimate goal of the method is to approximate any part of the solution manifold with a small number of basis functions $\{\phi_i \}_{i=1}^{N_{rb}}$. Thus, the reduced space is a linear approximation of the solution manifold as $\mathbb{V}_{fb}=$span$\{\phi_1,...,\phi_{N_{rb}}\}$.
The solution of the PDE is expressed as an expansion of the reduced basis functions $\phi_i(x)$ and the coefficients $a_i(s)$ in the Laplace domain,
\begin{equation}\label{RBexp}
p_{FOM}(x,s,\mu) \approx p_{ROM}(x,s,\mu) = \sum_{i=1}^{N_{rb}}\phi_{i}(x)a_i(s,\mu).
\end{equation}
POD is a standard way to generate basis functions and it is used for data compression and low-dimensional approximations and provides an orthogonal basis. The reduced basis is obtained by truncating the basis while keeping the essential information that ensures the desired accuracy of the results.
The generation of the basis consists of performing a sampling of the parameter space $\mathbb{P}_h \subset \mathbb{P}$ for $[\mu_1,...,\mu_{N_{\mu}}]$. The snapshots are collected into a snapshot matrix
\begin{equation}
\boldsymbol{S_N} = [p_{FOM}(\boldsymbol{x},s_1,\mu_1),...,p_{FOM}(\boldsymbol{x},s_{N_s},\mu_{N_{\mu}})],
\end{equation}
where $N_s$ is the number of complex frequencies. After solving  (\ref{Generalmatrix}), the snapshot matrix can be written as
\begin{equation}\label{SN1}
    \boldsymbol{S_N}=\begin{bmatrix} \boldsymbol{S_N}^{\sigma}\\ \boldsymbol{S_N}^{\gamma}\end{bmatrix} \in \mathbb{R}^{2N\times N_{s}},
\end{equation}
Note that the basis functions are orthogonal and can be generated by the singular value decomposition (SVD) of the snapshot matrix
\begin{equation}\label{eqsvd}
    \boldsymbol{S_N} = \mathbf{U}\mathbf{\Sigma}\mathbf{V}^T.
\end{equation}
Thus,  $\boldsymbol{\Phi} = [U_1,...,U_{N_{rb}}]\in \mathbb{C}^{2N\times N_{rb}}$, where $U_i$ corresponds to the $i$th singular vector. The singular values in $\mathbf{\Sigma}$, provides information about the reduction of the problem, defined as
\begin{equation}\label{EE0energy}
E/E_0 = \frac{\delta_i}{\sum_{i=1}^N \delta_i}, 
\end{equation}
where $\boldsymbol{\Sigma}=$diag$(\delta_i)$, $i=1,...,N_s$. In practice the number of basis funcions $N_{rb}$ can be chosen by a prescribed tolerance $\epsilon_{POD}$
\begin{equation}\label{romNrbselec}
    I(N_{rb}) = \frac{\sum_{i=1}^{N_{rb}} \delta^2_i}{\sum_{i=1}^{N} \delta^2_i} \geq 1-\hat{\epsilon}_{POD}.
\end{equation}
To preserve the structure of the high-fidelity matrix $\boldsymbol{K}$, a symplectic decomposition (PSD) with a symplectic Galerkin projection is used. Specifically, \textit{the cotangent-lift} method introduced in \cite{symplectic}, is applied. It ensures a symplectic matrix in block-diagonal form. Thus, the snapshot matrix (\ref{SN1}) is now considered in an extended form
\begin{equation}
\boldsymbol{S}_{\boldsymbol{N}cl}=\begin{bmatrix}\boldsymbol{S_N}^{\sigma},\boldsymbol{S_N}^{\gamma}\end{bmatrix}  \in \mathbb{R}^{N\times 2N_{s}},
\end{equation}
which can be decomposed in the same way as before by using an SVD to recover the corresponding POD basis, defined as  $\boldsymbol{\Phi} = [U_1,...,U_{N_{rb}}]\in \mathbb{C}^{N\times N_{rb}}$. Then, the symplectic basis is constructed as
\begin{equation}
   \boldsymbol{\Phi}_{cl} = \begin{bmatrix} \boldsymbol{\Phi} & \boldsymbol{0} \\ \boldsymbol{0} & \boldsymbol{\Phi} \end{bmatrix},
\end{equation}
where $\boldsymbol{\Phi}_{cl} \in \mathbb{C}^{2N\times 2N_{rb}}$.
As the ROM is built with a Galerkin projection, the solution can be represented as an expansion of POD basis functions that embed the spatial dynamics and the coordinate system in the Laplace domain.

The matrix expression for the real and imaginary parts of  (\ref{RBexp})  is
\begin{align}\label{bmatrix11}
&\begin{bmatrix} p_{ROM}^{\sigma}\\ p_{ROM}^{\gamma}\end{bmatrix} =  \begin{bmatrix} \Phi a^{\sigma} \\ \Phi a^{\gamma}\end{bmatrix}.
\end{align}
Inserting  (\ref{bmatrix11}) into (\ref{eqinarow}), the problem is similar to  (\ref{Generalmatrix}) but now consists of solving
\begin{gather}\label{matrixQrb}
  \mathbf{K}_{rb} \begin{bmatrix} a^{\sigma} \\ a^{\gamma}\end{bmatrix} = \mathbf{q}_{rb},
\end{gather}
where $ a^{\sigma}$ and $ a^{\gamma}$ are the real and imaginary part of the solution and $q_{rb}$ is the reduced right hand side function. Moreover $\mathbf{K}_{rb} = \boldsymbol{\Phi}_{cl}^T \mathbf{K}\boldsymbol{\Phi}_{cl}$, where
\begin{align}
  \mathbf{K}_{rb}  &= 
    \begin{bmatrix} \boldsymbol{\Phi}&0 \\ 0&\boldsymbol{\Phi}\end{bmatrix}^T
 \begin{bmatrix} \mathbf{K}^{\sigma} & -\mathbf{K}^{\gamma} \\ \mathbf{K}^{\gamma} & \mathbf{K}^{\sigma} \end{bmatrix} \begin{bmatrix} \boldsymbol{\Phi} & 0 \\ 0&\boldsymbol{\Phi}\end{bmatrix}, \\   \mathbf{q}_{rb} &= \boldsymbol{\Phi}_{cl}^T
  \begin{bmatrix} Q^{\sigma} \\ Q^{\gamma}\end{bmatrix}.
\end{align}
The operator can be written as
\begin{align}\label{GeneralMROM}
\mathbf{K}_{rb} =  \begin{bmatrix} \boldsymbol{\Phi}\boldsymbol{K}^{\sigma}\boldsymbol{\Phi}^T&-\boldsymbol{\Phi}^T\boldsymbol{K}^{\gamma}\boldsymbol{\Phi}\\\boldsymbol{\Phi}\boldsymbol{K}^{\gamma}\boldsymbol{\Phi}^T&\boldsymbol{\Phi}^T\boldsymbol{K}^{\sigma}\boldsymbol{\Phi} \end{bmatrix} = \begin{bmatrix} \boldsymbol{K}_{\Phi}^{\sigma} &-\boldsymbol{K}_{\Phi}^{\gamma}\\\boldsymbol{K}_{\Phi}^{\gamma}&\boldsymbol{K}_{\Phi}^{\sigma} \end{bmatrix},
\end{align}
where
\begin{subequations}\label{Krom}
\begin{align}\label{Krom1}
    &\boldsymbol{K}_{\Phi}^{\sigma} = (\sigma^2-\gamma^2)\mathcal{M}_{\Phi} + c^2\mathcal{S}_{\Phi}+\sigma\mathcal{B}_{\Phi}^{\sigma}, \\ &\mathcal{M}_{\Phi} = \boldsymbol{\Phi}^T\mathcal{M}\boldsymbol{\Phi},\\\label{Krom2}
   & \boldsymbol{K}_{\Phi}^{\gamma} = 2\sigma \gamma\mathcal{M}_{\Phi}+\gamma\mathcal{B}_{\Phi}^{\gamma}, \\ &\mathcal{S}_{\Phi} = \boldsymbol{\Phi}^T \mathcal{S}\boldsymbol{\Phi},\\\label{ROMfreqdepBCall}
    & \mathcal{B}_{\Phi}^{\sigma} = \Phi^T\mathcal{B}^{\sigma}\Phi,\\ &\mathcal{B}_{\Phi}^{\gamma} = \Phi^T\mathcal{B}^{\gamma}\Phi.
\end{align}
\end{subequations}
For the case where the system is excited with an initial condition the expression becomes
\begin{align}
  \mathbf{q}_{rb}&=   \begin{bmatrix} \boldsymbol{\Phi}&0 \\ 0&\boldsymbol{\Phi}\end{bmatrix}^T
 \begin{bmatrix} Q^{\sigma}\\ Q^{\gamma}\end{bmatrix} =\begin{bmatrix} Q^{\sigma}_{\Phi}\\ Q^{\gamma}_{\Phi}\end{bmatrix},
\end{align}
where
\begin{subequations}
\begin{align}
    &Q^{\sigma}_{\Phi} = \sigma \hat{p}_0\mathcal{M}_{\Phi q}, &\quad Q^{\gamma}_{\Phi} = \gamma \hat{p}_0\mathcal{M}_{\Phi q},\\
    &\mathcal{M}_{\Phi q} = \boldsymbol{\Phi}^T\mathcal{M}.
\end{align}
\end{subequations}
The solution $[p_{rb}^{\sigma}, p_{rb}^{\gamma}]^T$ is finally transformed to the time-domain by applying Weeks method.\\

For frequency-independent boundary conditions, the parameter  $\mu$  corresponds to the surface impedance $Z_s$. It is sampled to generate the snapshots for every complex frequency so that $\boldsymbol{S_N}\in \mathbb{R}^{N\times h2N_{s}}$, where $h$ is the number of sampled impedance values.\\

The frequency-dependent boundary conditions are treated in a similar way. First, the real and imaginary parts of the pressure are computed from (\ref{hardwalls}). These values are used in (\ref{verloc}), to compute the real and imaginary components of the normal velocity at the boundary. Making use of (\ref{BCImpedaceReal}), the admittance at the boundaries is computed. Finally, (\ref{matrixQrb}) is solved.\\

To compare the performance of the ROM against the FOM, the speedup is introduced
\begin{equation}
    speedup=\frac{CPU_{FOM}}{CPU_{ROM}},
\end{equation}
where $CPU_{FOM}$ is the time needed to solve the FOM for a single parameter value and $CPU_{ROM}$ the time needed to solve the ROM a single parameter value with the selected number of basis $N_{rb}$. Note that the speedup measures only the online stage.


\section{Numerical results}
We now examine numerical test cases to offer insights into the different properties of the FOM and ROM. 

\subsection{Full Order Model verification}

\begin{figure}[ht!]
\begin{subfigure}{1\textwidth}
    \centering
    \includegraphics[width = 0.7\linewidth]{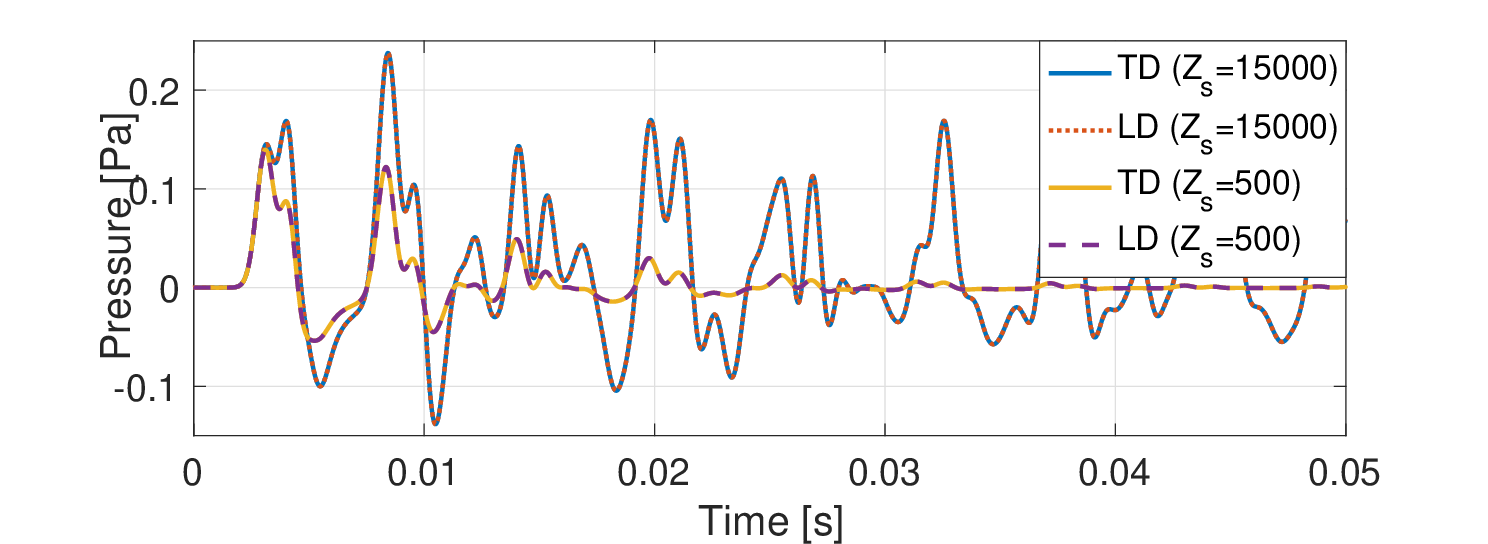}
    \caption{Frequency-independent boundaries.}
    \label{fig:freq_indpSOL}
\end{subfigure} %
\begin{subfigure}{1\textwidth}
    \centering
    \includegraphics[width = 0.7\linewidth]{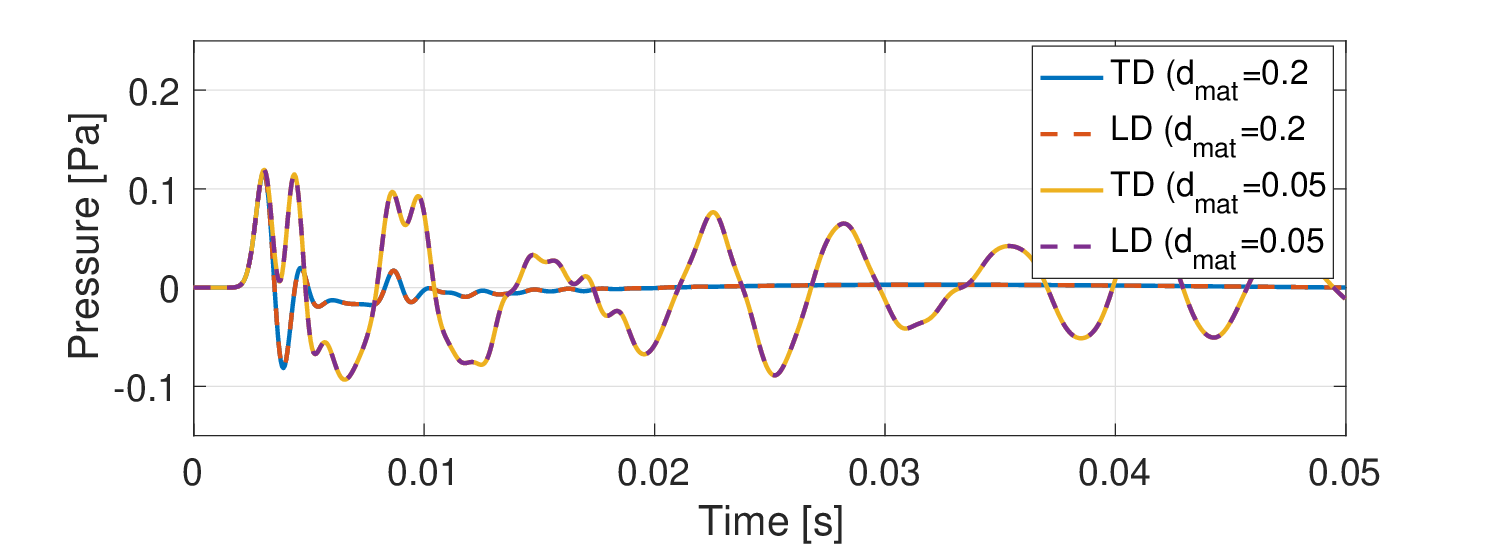}
    \caption{Frequency-dependent boundaries.}
    \label{fig:freq_dpSOL}
\end{subfigure} %
\begin{subfigure}{1\textwidth}
    \centering
    \includegraphics[width = 0.7\linewidth]{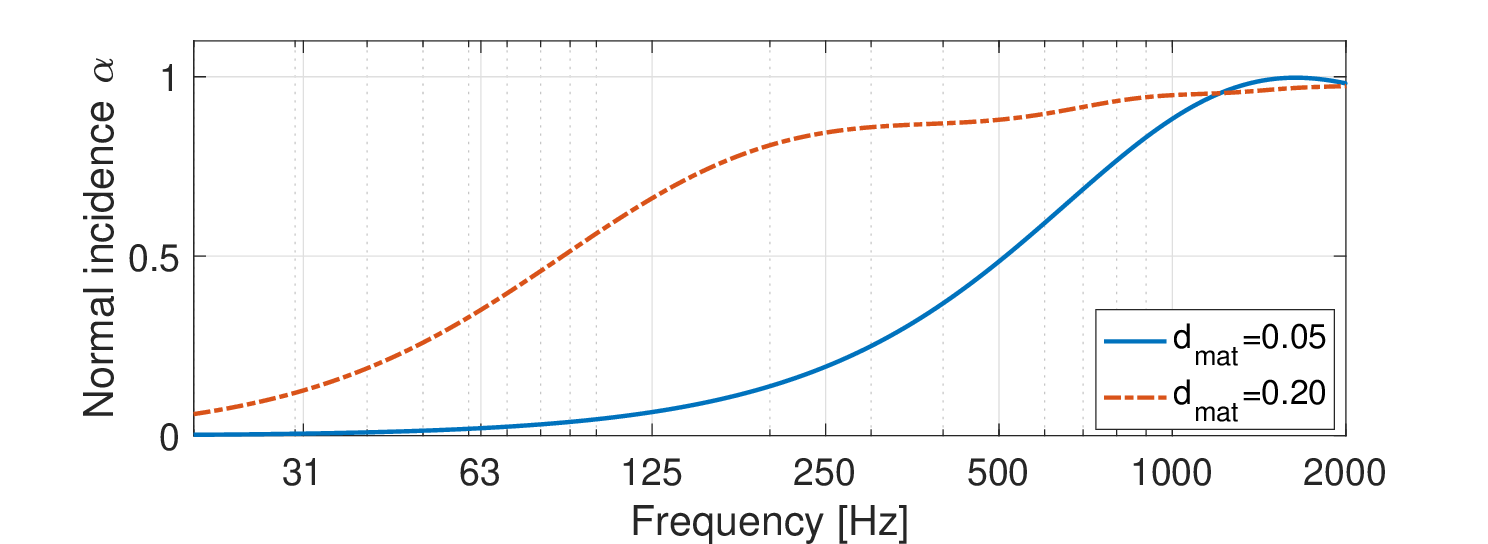}
    \caption{Absorption coefficient for different material thickness $d_{mat}$ (in meters)}
    \label{fig:abscoef1}
\end{subfigure} %

\caption{FOM impulse response simulations for different parameter values and absorption coefficients. The source location is $(s_x,s_y) = (1, 1)$ m and the receiver location is $(r_x,r_y)=(0.2, 0.2)$ m. Simulations were carried out with $P=4$, $\Delta t = 5.9\times 10^{-6}$ s, $N_e=15$ and $N_e=20$, $N_s= 3000$ and $(\sigma, b)=(10, 1000)$.}
\end{figure}

The Laplace domain SEM for room acoustics is verified via comparisons with a time domain SEM solver \cite{SEM}.
First, a 2D domain ($2$ m $\times 2$ m) with frequency-independent boundary conditions is considered. The number of elements per Cartesian direction is set to $N_e=20$ and a basis order of $P=4$ is used ($N=6561$). The spatial resolution in this case corresponds to roughly 13 points per wavelength ($PPW$) at 1 kHz. The model is excited with a Gaussian pulse as initial condition (\ref{initial_cond}) with $\sigma_g=0.2$ at $(s_x,s_y)=(1,1)$ m.
First, the model is tested with frequency-independent boundaries with two surface impedances of $Z_s=[500\quad15000]$ kgs$^{-1}$m$^{-2}$, which correspond to the normal incidence absorption coefficient value $\alpha_{norm}=[0.99\quad0.10]$.  The time step was selected following the  global Courant-Friedrichs-Lewy (CFL) condition for the time-domain solver defined as \cite{SEM}. The parameters for the Weeks time reconstruction when using $N_s =3000$ are $\sigma=10$ and $b=1000$. Figure \ref{fig:freq_indpSOL} shows the impulse responses at $(r_x,r_y)=(0.2, 0.2)$ showing good agreement between the two solvers. For the particular case of $Z_s=15000$ kgs$^{-1}$m$^{-2}$, the error at the receiver point at $t=0.1$ s is $\epsilon=|p_{\text{Time}}-p_{\text{Laplace}}|=8\times 10^{-5}$ Pa.

 Secondly, frequency-dependent boundaries are considered for the same domain, source and receiver positions and simulation parameters, except that $N_e=15$ ($N=3721$). The boundary is modelled as a porous material mounted on a rigid backing. The flow resistivity of the material is $\sigma_{mat} = 10 000$ Nsm$^{-4}$ and the thickness as $d_{mat}=[0.05\quad 0.2]$ m. The surface impedance are estimated using Miki's model \cite{Mikis} in conjunction with a transfer matrix method \cite{allardbook}. The corresponding absorption coefficient (\ref{abscoeff}) is presented in Figure \ref{fig:abscoef1}. Figure \ref{fig:freq_dpSOL} confirms a good match between the two solvers.  For the particular case of $d_{mat}=0.05$ m, the error was calculated at the receiver point and $t=0.05$ s, as $\epsilon=|p_{\text{Time}}-p_{\text{Laplace}}|=3\times 10^{-5}$ Pa. 


\subsubsection{Influence of the Weeks parameters}

\begin{figure}[ht!]
\begin{subfigure}{0.5\textwidth}
    \centering
    \includegraphics[width = 1\linewidth]{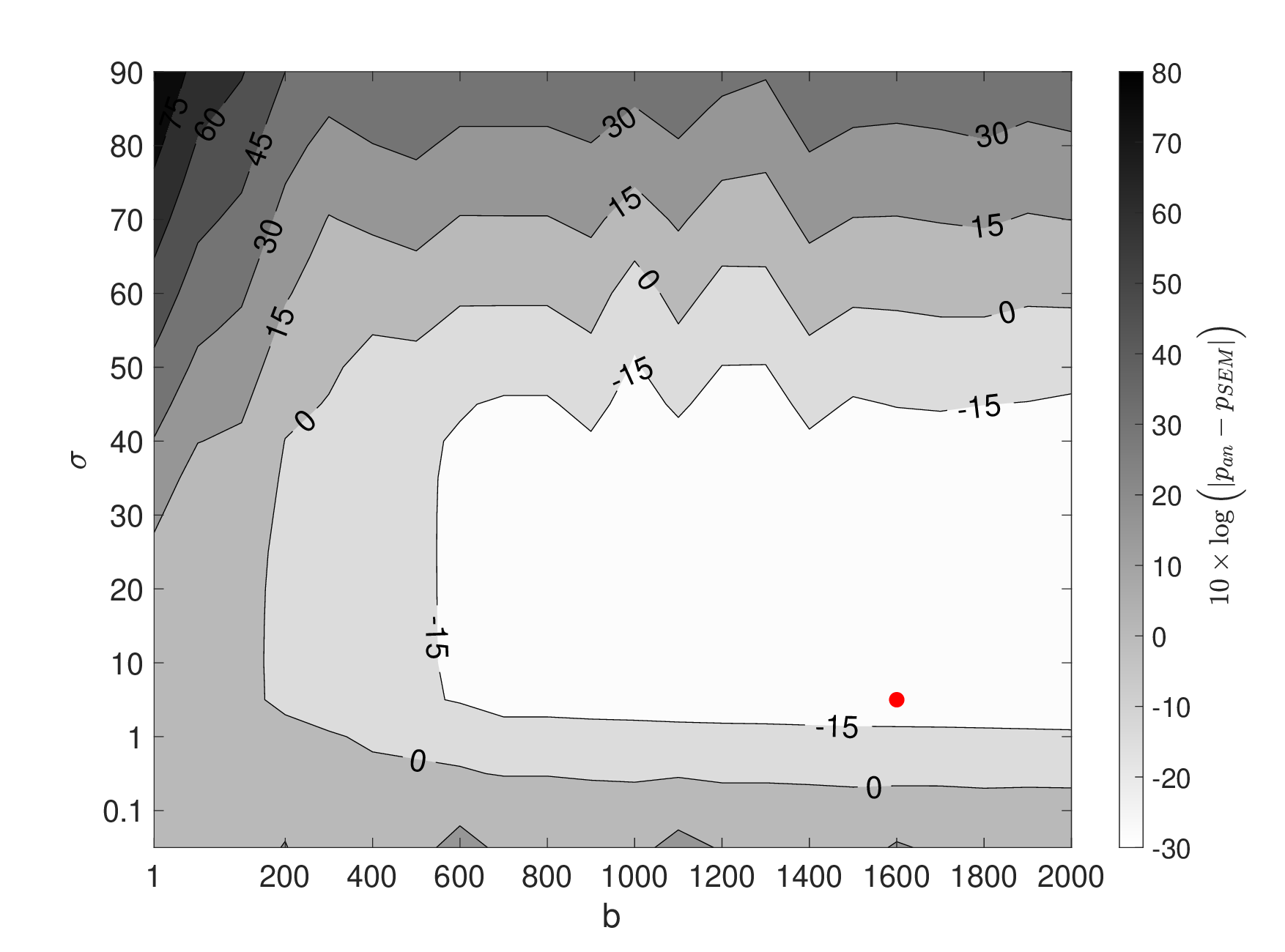}
    \caption{Rigid boundaries with $N_s= 7000$.}
    \label{sfigrbc:3}
\end{subfigure} %
\begin{subfigure}{0.5\textwidth}
    \centering
    \includegraphics[width = 1\linewidth]{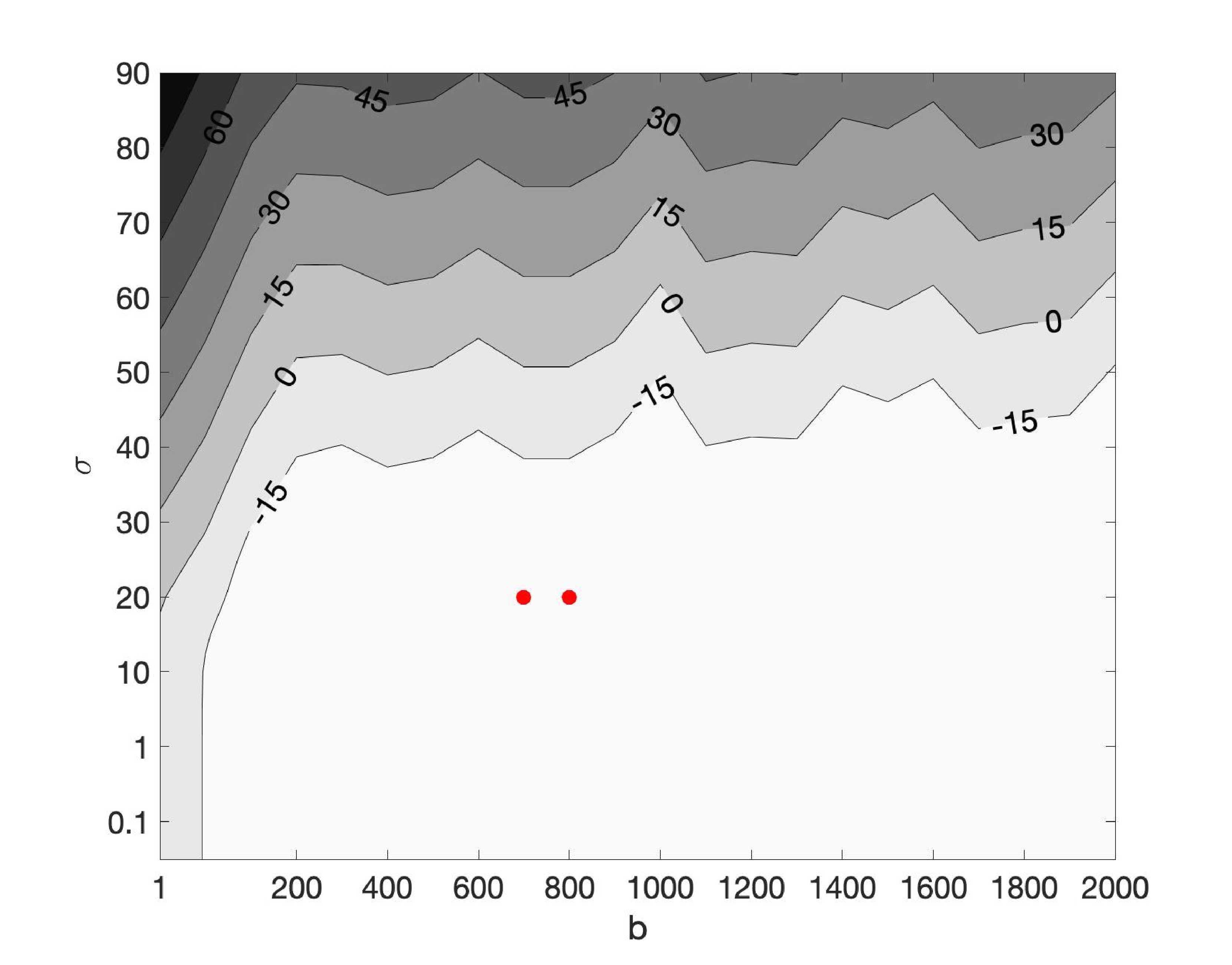}
    \caption{$Z_s = 2000$ kgs$^{-1}$m$^{-2}$ with $N_s= 7000$.}
    \label{sfigz2000:3}
\end{subfigure} %

\caption{Contour plot of the error obtained using 11 equally spaced points for $\sigma \in[0.1, 2000]$ and 11 equally spaced points for $b \in[0.1,90]$ for three different boundary conditions to obtain (\ref{optimization}).}
\label{fig:parameters}
\end{figure}

The free parameters of the Weeks method ($\sigma$, $b$) are identified using (\ref{optimization}). The error is estimated using $N_s = 7000$ and $11$ equally spaced points for $\sigma \in[0.1, 2000]$ and $11$ equally spaced points for $b \in[0.1,90]$ for rigid boundary conditions and frequency-independent boundaries with $Z_s=2000$ kgs$^{-1}$m$^{-2}$. Figure \ref{fig:parameters} shows the results and the pair of parameter values that provide the lowest error are marked with a red dot. Contour lines present the error in $dB$ given by $10\log_{10}(|p_{Time}-p_{Laplace}|)$. The parameters can be selected for those values that lie inside the region where the error is the minimal. For the rigid boundary case this region is smaller, whereas it is found to be larger once some absorption is added at the boundary. As $Z_s$ decreases, i.e., increasing the absorption at the boundaries, the region becomes larger. This finding is good news from an acoustical point of view, as in most cases some absorption is present at the boundary, making the choice of the two free parameters less crucial.

\subsection{Reduced Order Model}\label{sectionROM}
In this section, the reduced basis method is applied to different test cases to illustrate the potential of the method. The varying parameter of the system is the absorption properties of the boundaries. 

\subsubsection{2D ROM with frequency-independent boundaries}
The  frequency-independent test case as before is now considered with $N=6561$, which for a six-elements per wavelength thumb rule \cite{sixelem} the upper frequency is $2$ kHz. The ROM is constructed by generating snapshots when sampling the surface impedance values uniformly $Z_s = [500, 1500, 2500, ..., 15500]$ kgs$^{-1}$m$^{-2}$. 

\begin{figure}[h!]

\begin{subfigure}{0.5\textwidth}
    \centering
    \includegraphics[width = 1\linewidth]{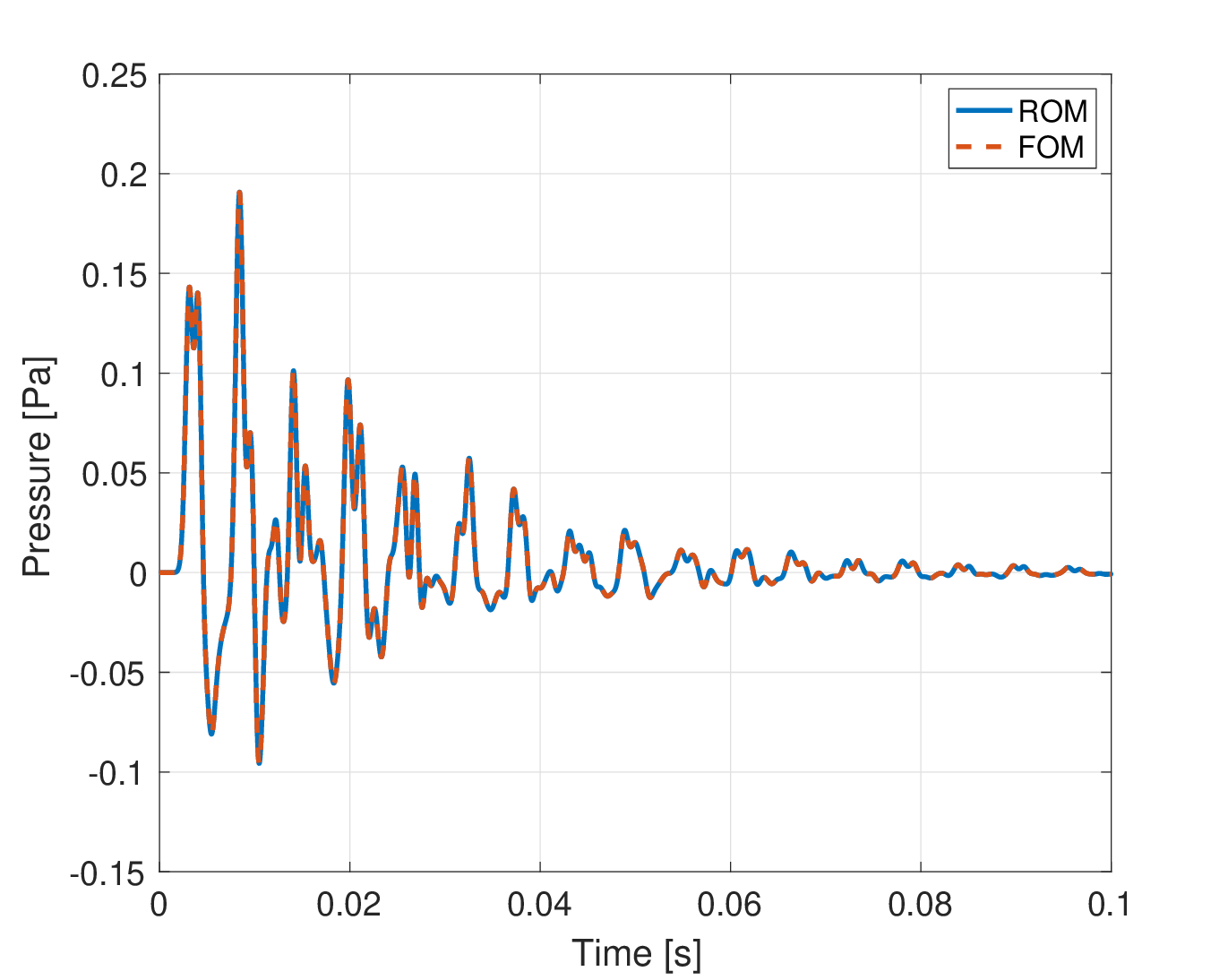}
    \caption{Frequency-independent boundaries with $Z_s=5000$ kgs$^{-1}$m$^{-2}$ and  $N_{rb}=300$.}
    \label{fig:ROM_freIND1}
\end{subfigure}
\begin{subfigure}{0.5\textwidth}
    \centering
    \includegraphics[width = 1\linewidth]{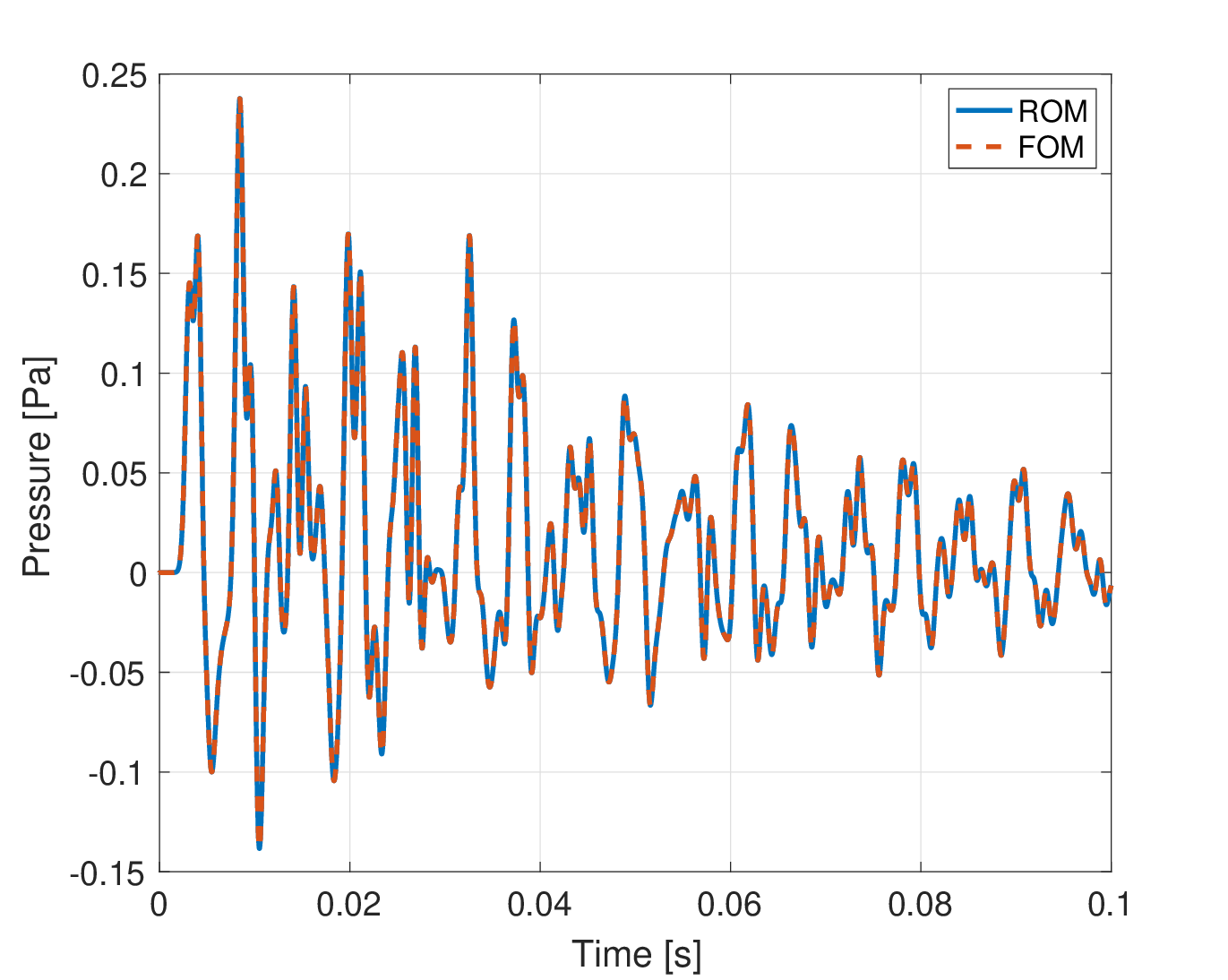}
    \caption{Frequency-independent boundaries with $Z_s=15000$ kgs$^{-1}$m$^{-2}$  and $N_{rb}=300$.}
    \label{fig:ROM_freIND2}
\end{subfigure}
\begin{subfigure}{0.5\textwidth}
    \centering
    \includegraphics[width = 1\linewidth]{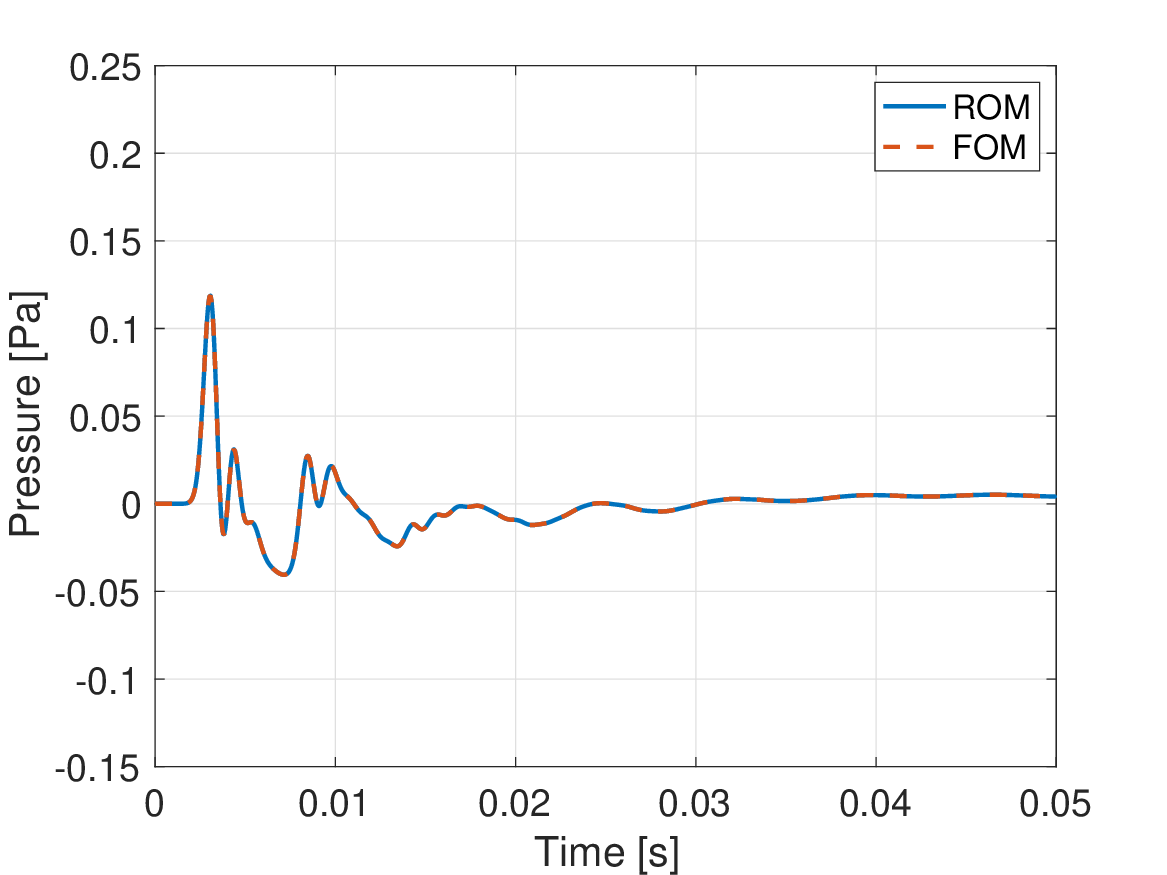}
    \caption{Frequency-dependent boundaries with $d=0.15$ m and $N_{rb}=150$.}
    \label{fig:ROM_freDep1}
\end{subfigure}
\begin{subfigure}{0.5\textwidth}
    \centering
    \includegraphics[width = 1\linewidth]{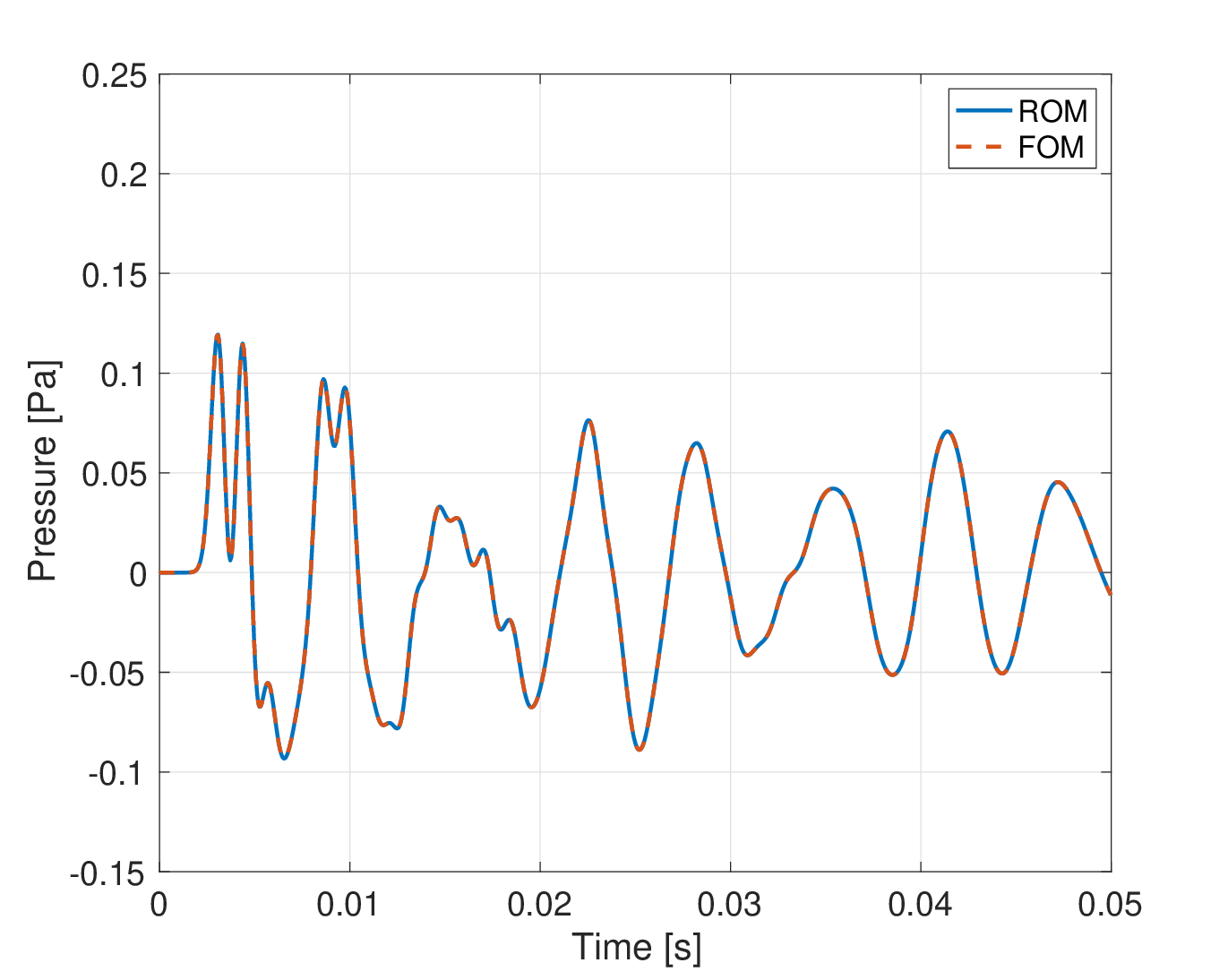}
    \caption{Frequency-dependent boundaries with $d=0.05$ m and $N_{rb}=150$.}
    \label{fig:ROM_freDep2}
\end{subfigure}
\caption{Simulated pressure using the 2D FOM and ROM for different parameter values $Z_s$ and $d_{mat}$. }
\label{fig:ROM_fre}

\end{figure}

Figure \ref{fig:ROM_freIND1} and Figure \ref{fig:ROM_freIND2} show the simulation results with surface impedances of 5000 kgs$^{-1}$m$^{-2}$ and 15000 kgs$^{-1}$m$^{-2}$, that were not included in the snapshot matrix. The corresponding FOM simulations are used for verification. The number of basis functions ($N_{rb}=300$) were selected using (\ref{romNrbselec}) for $\hat{\epsilon}_{POD} = 10^{-6}$. Results show a good match between the ROM and FOM with an absolute error, calculated at the receiver point at $t=0.1$ s and for $Z_s=15000$ kgs$^{-1}$m$^{-2}$, of $\epsilon=|p_{\text{FOM}}-p_{\text{ROM}}|=4.3\times 10^{-9}$ Pa. The relative error for this particular case is given by $\epsilon_{rel}=|\frac{p_{\text{FOM}}-p_{\text{ROM}}}{p_{\text{FOM}}}|=6.5\times 10^{-7}$.

\begin{figure}[h!]
    \centering
    \includegraphics[width = 0.5\linewidth]{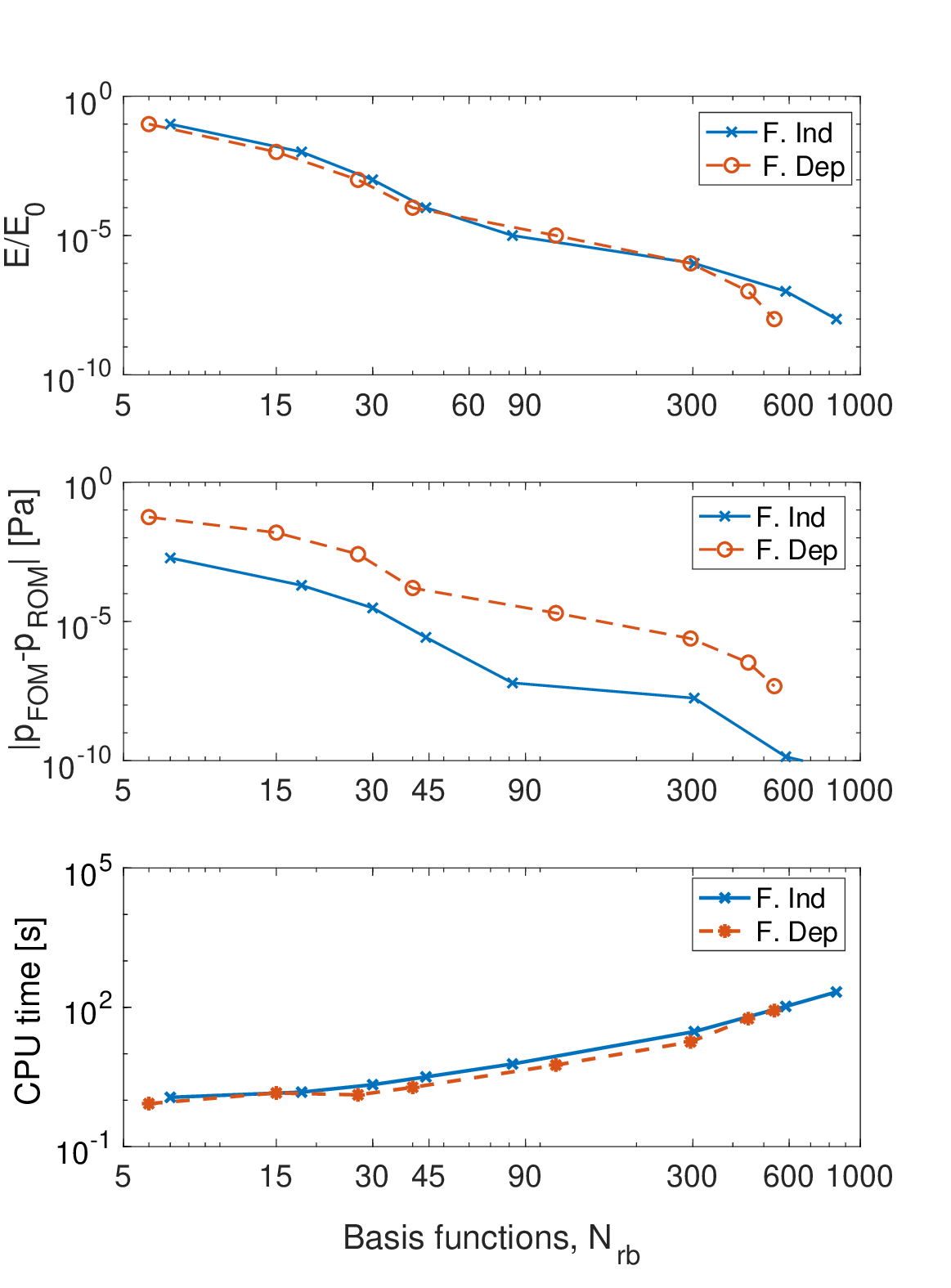}
\caption{ Error and cpu time of the 2D ROM for different basis functions. Simulations were carried out with a fixed $PPW=13$ for frequency-independent boundaries and $PPW=10$ for frequency-independent at $1000$ Hz,  $P=4$, $Z_s = 5000$  kgs$^{-1}$m$^{-2}$ and $d_{mat}=0.15$ m.}
\label{fig:ROMERROR}
\end{figure}

\begin{figure}[h!]

\begin{subfigure}{0.5\textwidth}
    \centering
    \includegraphics[width = 1\linewidth]{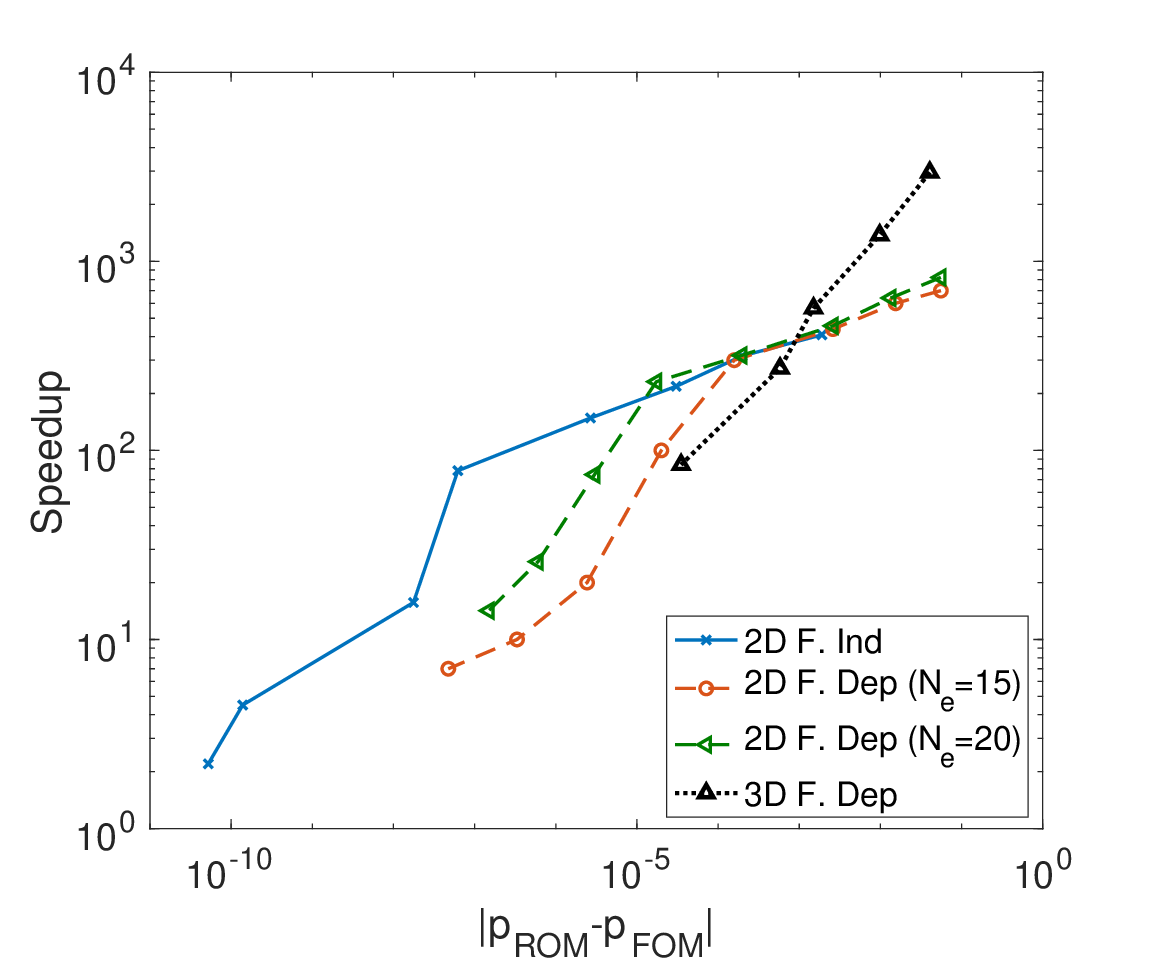}
    \caption{Speedup as a function of error.}
    \label{fig:ROMSPEEDUP_a}
\end{subfigure} %
\begin{subfigure}{0.5\textwidth}
    \centering
    \includegraphics[width = 1\linewidth]{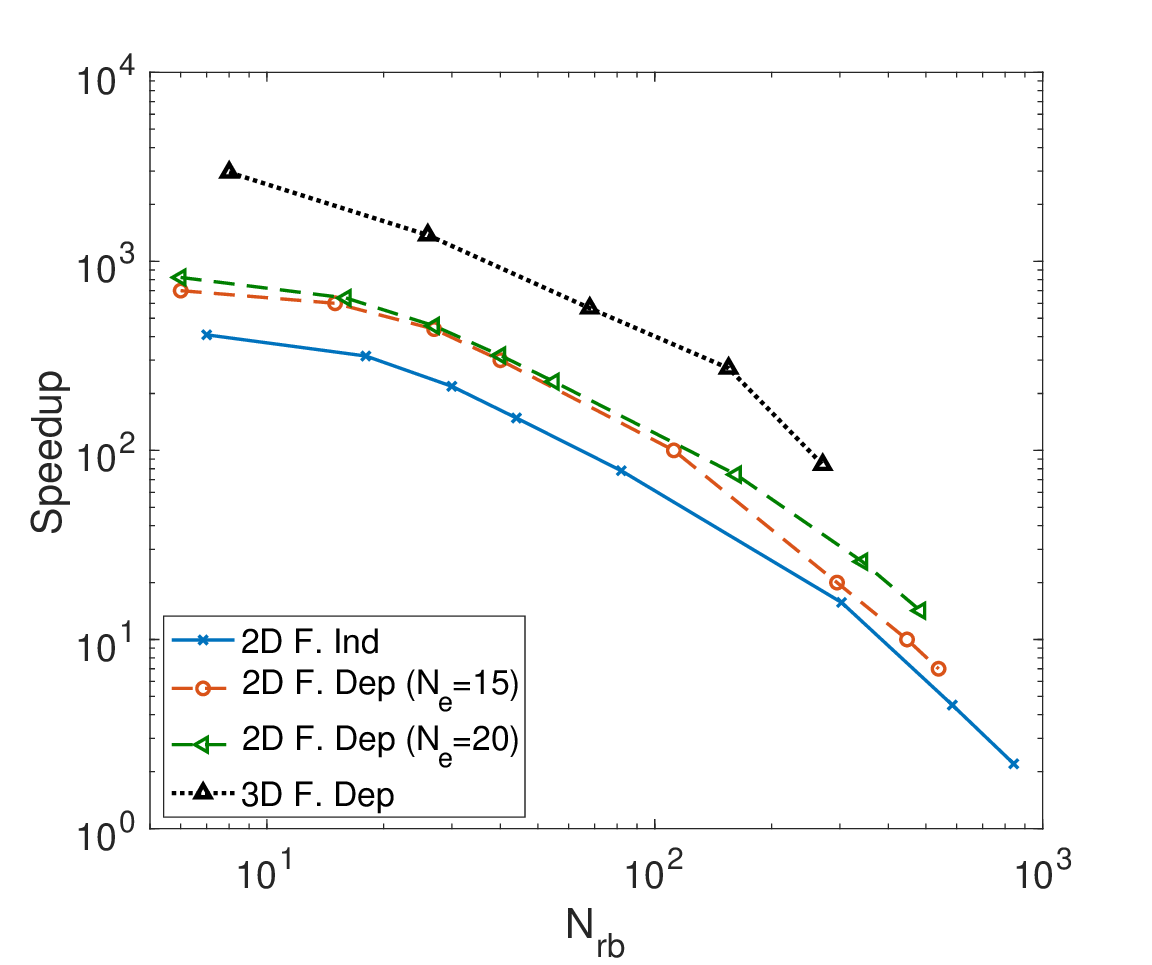}
    \caption{Speedup as a function of $N_{rb}$.}
    \label{fig:ROMSPEEDUP_b}
\end{subfigure}
\caption{\label{fig:ROMSPEEDUP}{Speedup for 2D and 3D. (a) Speedup as function of error, (b) Speedup as function of $N_{rb}$.}}
\end{figure}

The speedup and accuracy of the ROM are investigated by running simulations for different $N_{rb} = [7, 18, 30, 44, 82, 303, 585, 842]$. The singular energy distribution, error, and computational cost are presented as a function of the number of basis functions in  Figure \ref{fig:ROMERROR}. The speedup for different number of basis and error values is shown in Figure \ref{fig:ROMSPEEDUP}.
The convergence for the ROM Laplace solver is demonstrated by looking at the errors as a function of $N_{rb}$. As expected, including more basis functions to the ROM improves the accuracy but reduces the speedup. The speedup is about 400 when the error is $2 \times 10^{-3}$ Pa, and 200 for an error of $3\times10^{-5}$ Pa. 

\subsubsection{2D ROM with frequency-dependent boundaries}
The  frequency-dependent test case is considered again. The thickness of the modeled material is uniformly sampled to generate the snapshots as $d_{mat}=[ 0.02, 0.12, 0.22]$ m.


The results are presented in Figure \ref{fig:ROM_freDep1} and Figure \ref{fig:ROM_freDep2} for values that were not sampled to construct the ROM and compared with the corresponding FOM high-fidelity solution for verification purposes. The number of basis functions ($N_{rb}=150$) were selected for $\hat{\epsilon}_{POD}= 10^{-6}$. A good match between the ROM and FOM shown in Figure \ref{fig:ROM_freDep1} and Figure \ref{fig:ROM_freDep2}. The absolute error, calculated at the receiver point at $t=0.05$ s and for $d_{mat}=0.05$ m, is $\epsilon=|p_{\text{FOM}}-p_{\text{ROM}}|=2\times 10^{-9}$ Pa. Moreover, the relative error for this particular case is $\epsilon_{rel}=|\frac{p_{\text{FOM}}-p_{\text{ROM}}}{p_{\text{FOM}}}|=3.6\times 10^{-7}$.


The computational cost and speedup for different number of basis and error values are presented in Figure \ref{fig:ROMERROR} and Figure \ref{fig:ROMSPEEDUP}, and are similar to the frequency-independent case. When $N_e=15$ ($N=3725$), the speedup is about 700 when the error is $5.5\times10^{-2}$ Pa, and reduces to 100 for an error of $2\times10^{-5}$ Pa. A second ROM with $N_e=20$ ($N=6561$) is constructed to compare the effect of increasing the DOF. Figure \ref{fig:ROMSPEEDUP} shows that the ROM constructed with $N_e=20$ results in to higher speedups than the ROM with $N_e=15$. Note that Figure \ref{fig:ROMSPEEDUP} shows different speedup values for the frequency-independent and frequency-dependent cases. The reason is that the computational time of the FOM is higher for the frequency-dependent case, which leads into different speedup values. 


\subsubsection{3D cube ROM with frequency-dependent boundaries}
\begin{figure}[h]
    \centering
\includegraphics[width=0.8\linewidth]{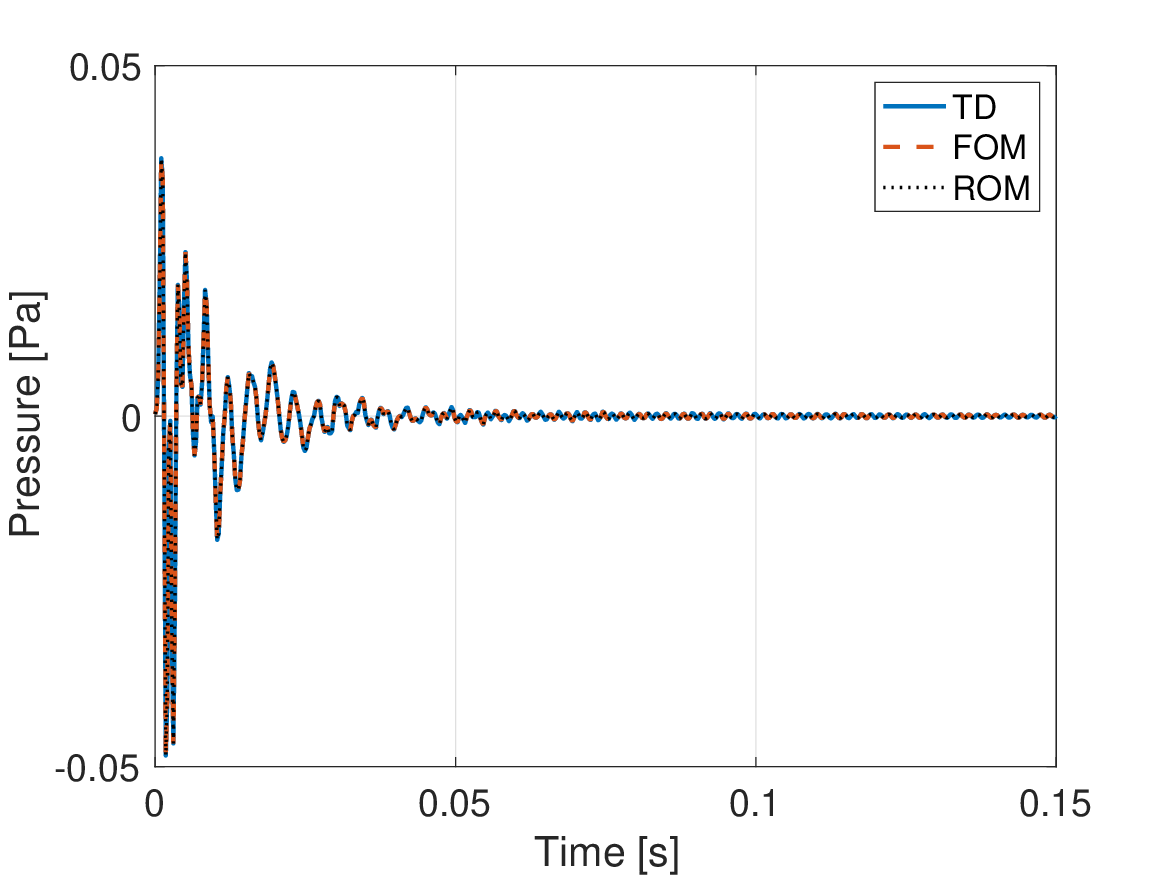}
\caption{\label{fig:3dtdld}{3D TD, LD FOM and ROM impulse response simulations. The source location is $(s_x,s_y,s_z)=(0.5, 0.5, 0.5)$ m ad the receiver location is $(r_x, r_y, r_z)=(0.25, 0.1, 0.8)$ m. Simulations were carried out with $P=4$, $N_e=8$, $N_s=1800$, $(\sigma,b)=(20, 800)$ and $N_{rb}=155$.}}

\end{figure}

This section considers a 3D $1$ m $\times 1$ m $\times 1$ m cube shaped room. A ROM with frequency-dependent boundaries is constructed. The boundaries are modeled using Miki's model \cite{Mikis}, with a flow resistivity of $\sigma_{mat} = 10 000$ Nsm$^{-4}$, where the material thickness $d_{mat}=[ 0.02, 0.12, 0.22]$ m is parametrized to create the ROM. The simulations are carried out using a number of elements per direction of $N_e=8$ and $P=4$ ($N=35937$). Assuming a fixed number of points per wavelength of $PPW=10$, the upper frequency limit is 1 kHz. The initial condition is a Gaussian pulse with $\sigma_g=0.2$ m$^2$. The source is placed at $(s_x, s_y, _z)=(0.5, 0.5, 0.5)$ m and the receiver is place at $(r_x, r_y, r_z)=(0.25, 0.1, 0.8)$  m. For this case, the number of frequencies is chosen to be $N_s=1800$ and the Week parameters are $\sigma=20$ and  $b=800$. Figure \ref{fig:3dtdld} shows the verification of the 3D solver against a time domain solver \cite{SEM} and the resulted ROM for $d_{mat}=0.05$ m and $N_{rb}=155$. For this particular case, the error between the FOM and the ROM is $5.8\times 10^{-4}$ Pa.\\
During the online phase of the RBM, different ROMs are built to compare the error and the speedup against the FOM. Figure \ref{fig:ROMSPEEDUP} shows the speedup of the ROM for different error values compared with the 2D cases. Results show a speedup between two and three orders of magnitude depending on the error tolerance chosen. The speedup is more than 1000 when the error tolerance is almost $10^{-2}$, and reduces to 100 for an error tolerance of $10^{-5}$. Note that the error for the 3D case in Figure \ref{fig:ROMSPEEDUP_a} is larger compared to the 2D cases due to a lower number of elements per direction $N_e$ and thus, a lower resolution. A better comparison is shown in Figure \ref{fig:ROMSPEEDUP_b} where the speedup is higher for the 3D case for the same number of basis.

\begin{figure}[h]
    \centering
    \includegraphics[width = 0.6\linewidth]{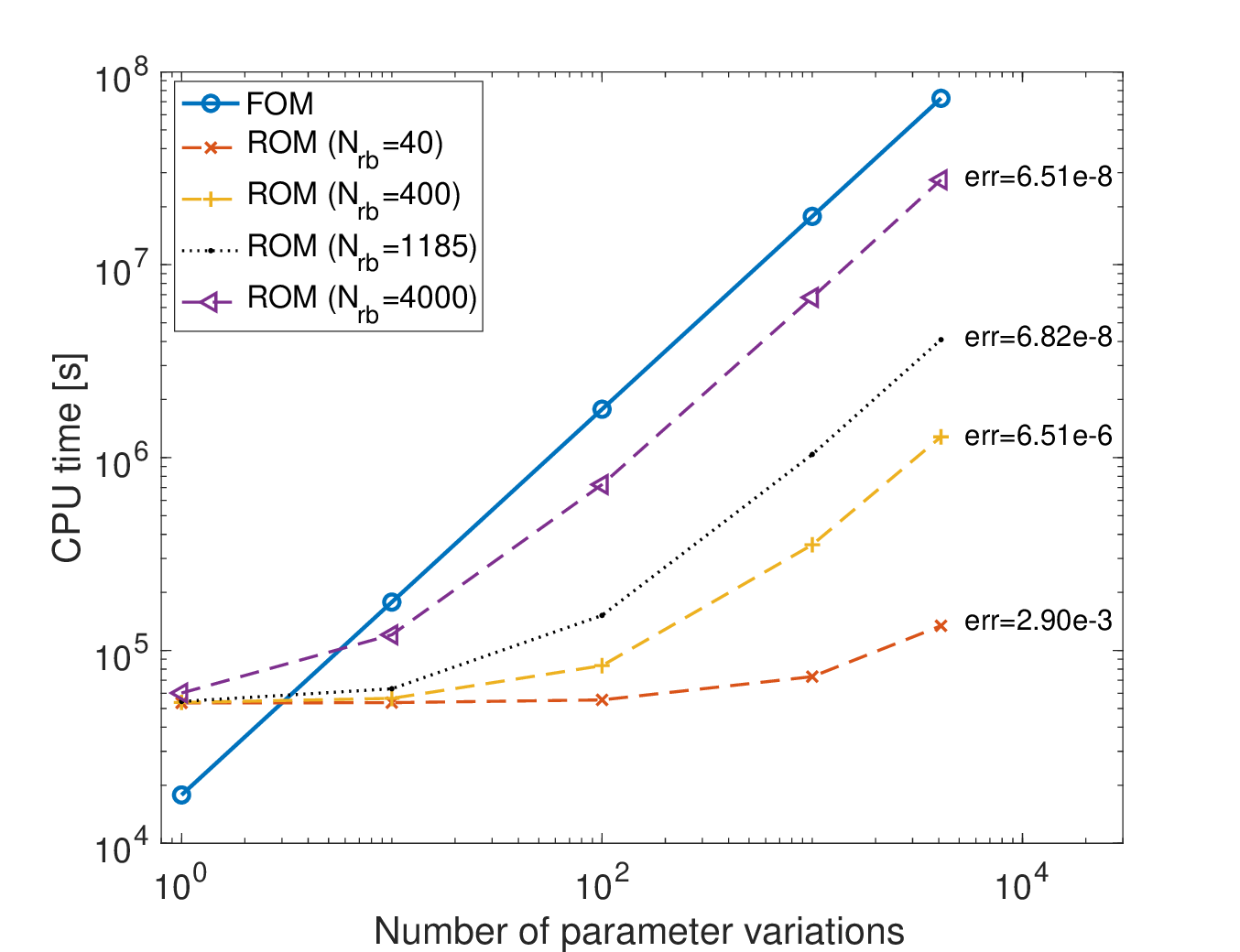}
    \caption{Computational cost of ROM including offline stage.}
    \label{fig:offlinecost1}
\end{figure}

The ROM constructed with three FOM simulations for different values of $d_{mat}$, requires 8GB of storage. The performance of the ROM including the offline stage cost is analyzed. The ROM is constructed with three FOM simulations for different values of $d_{mat}$. Thus, the computational time of the ROM begin after the time taken by the three FOM simulations. Figure \ref{fig:offlinecost1} shows the computational cost of the FOM and ROM with different $N_{rb}$. As an example, four different parameter values, $d_{mat}$, per surface need to be simulated, which leads to a total number of $4096$ simulations to explore all the possibilities. For the case where $N_{rb}=40$, the ROM is three orders of magnitude faster than the FOM with an error equal to $2.9\times 10^{-3}$ Pa. A singular energy decay analysis revealed that most energy is concentrated within the first $1185$ basis function, so the reduced model with $N_{rb}>1185$ does not significantly improve the error as shown in Figure \ref{fig:offlinecost1}). Thus, in this case, the interest is to construct a ROM with $N_{rb}< 1185$ and probably much smaller. Note that the offline stage consist of FOM simulations that are independent to each other. Thus, the computational time of this stage can be reduced, e.g., computing the FOM for different parameter values in parallel.

\subsubsection{Effects on the singular values decay and number of sampled parameters}
\begin{figure}[h!]
\begin{subfigure}{0.5\textwidth}
    \centering
    \includegraphics[width = 1\linewidth]{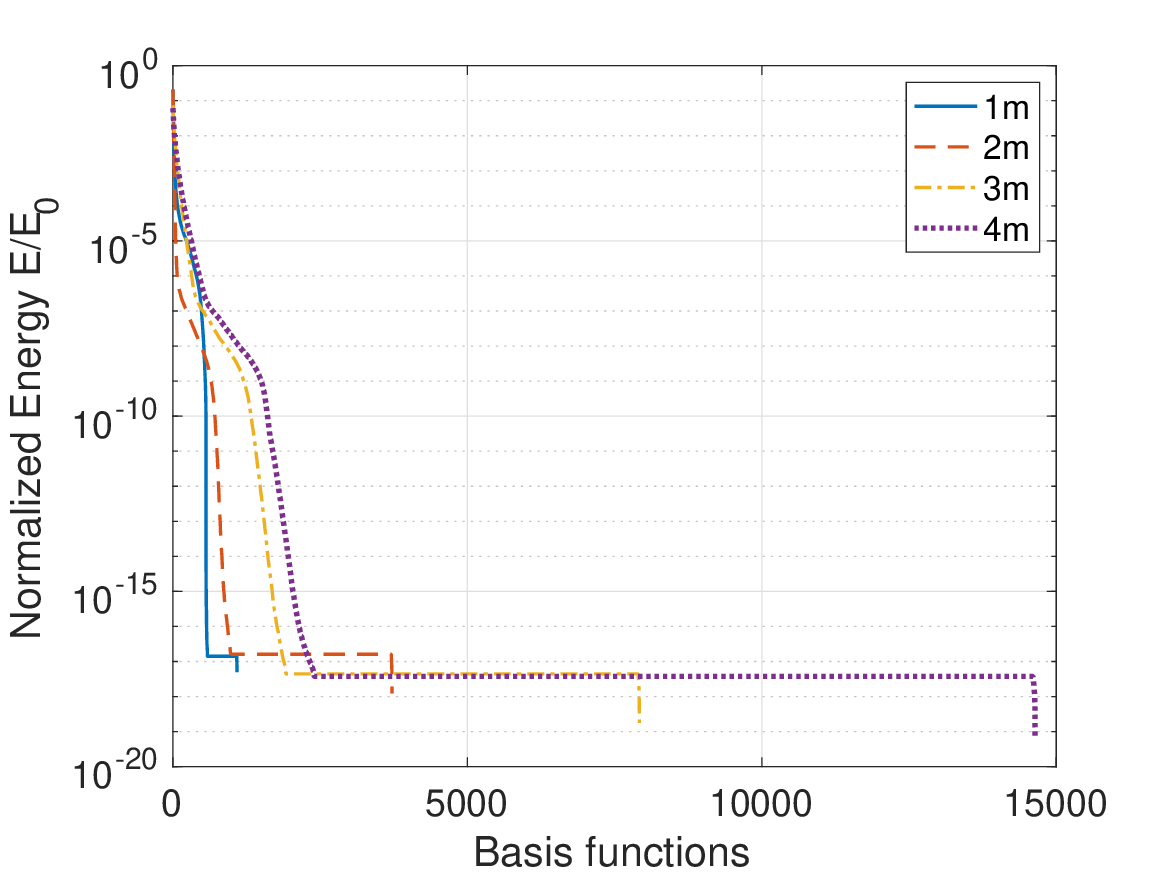}
    \caption{Energy decay for different domain side lengths with a fixed $PPW=10$ at $1000$ Hz.}
    \label{sfig:EnergtdecayAnaly1}
\end{subfigure} %
\begin{subfigure}{0.5\textwidth}
    \centering
    \includegraphics[width = 1\linewidth]{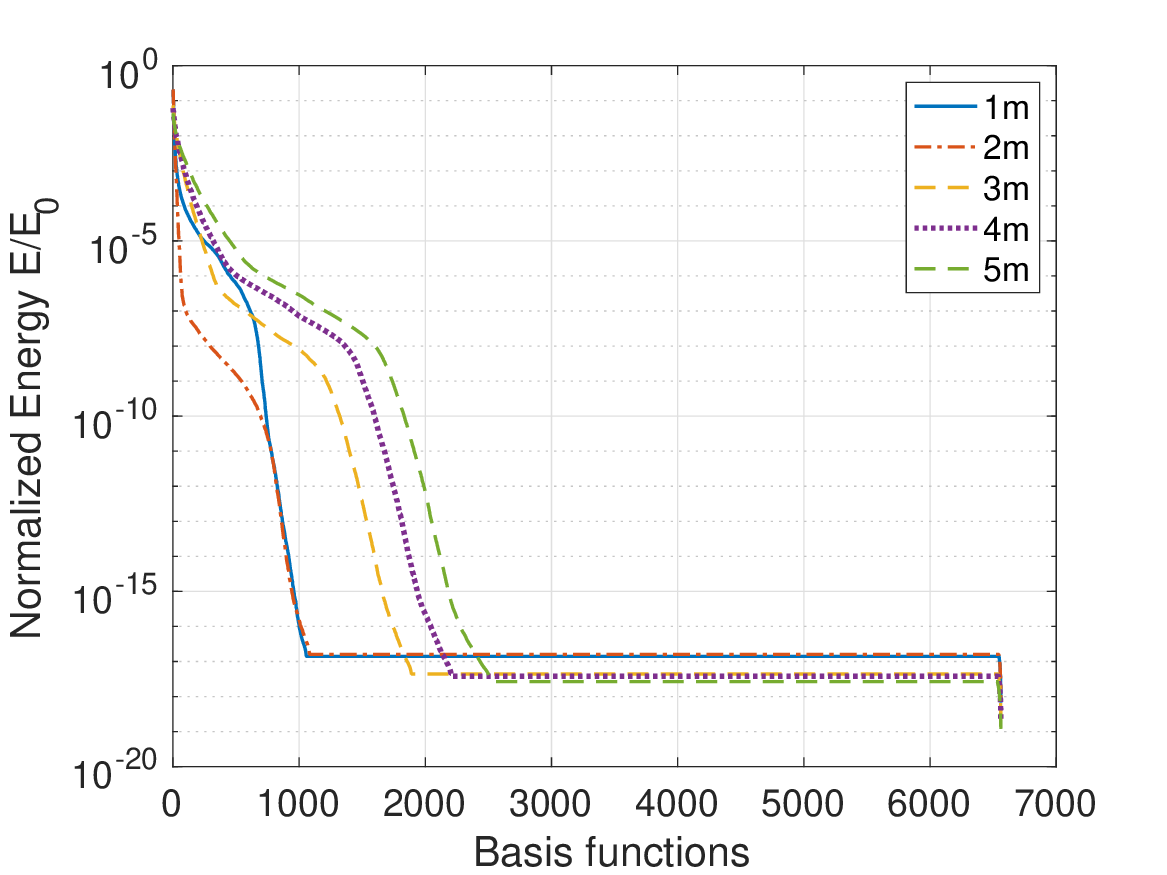}
    \caption{Energy decay for different domain lengths with a fixed number of elements $N_e = 20$.}
    \label{sfig:EnergtdecayAnaly2}
\end{subfigure}
\begin{subfigure}{0.5\textwidth}
    \centering
    \includegraphics[width = 1\linewidth]{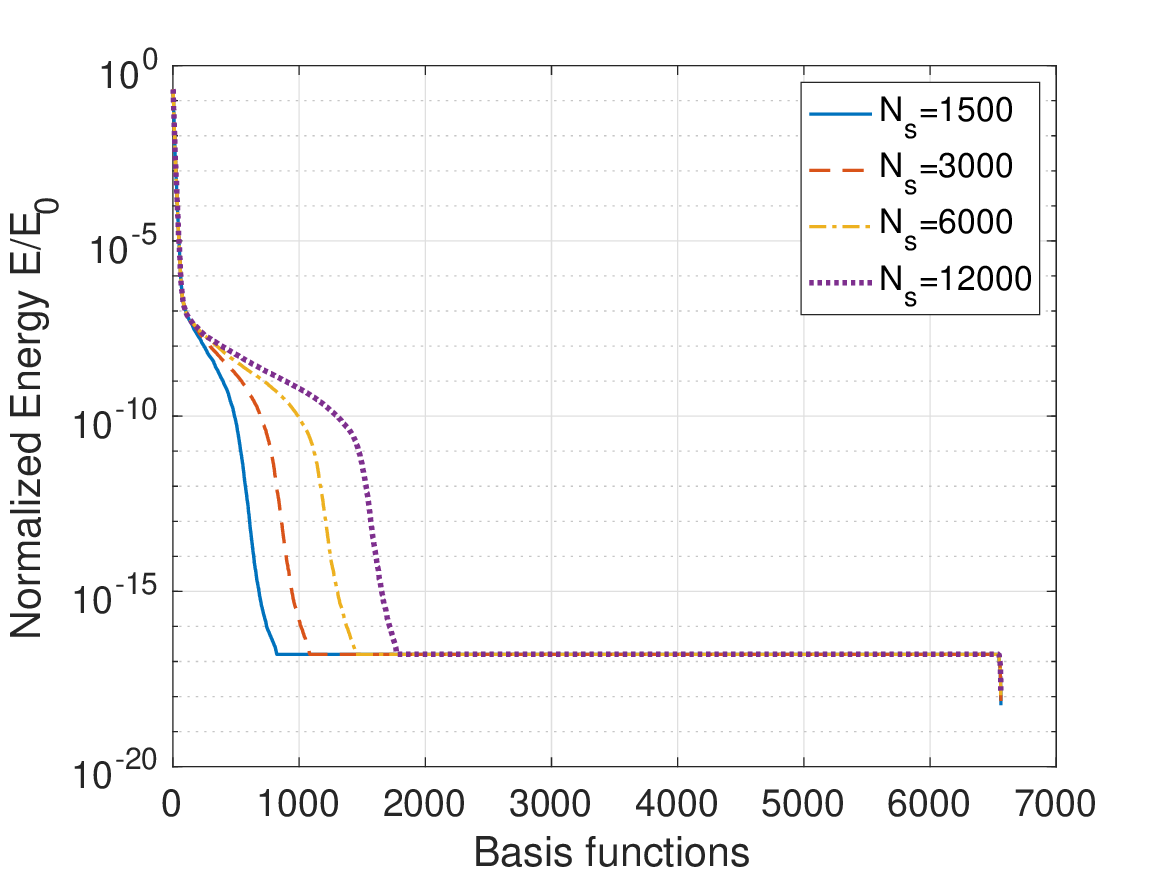}
    \caption{Energy decay  for different number of complex frequencies $N_s$.}
    \label{sfig:EnergtdecayAnaly4}
\end{subfigure}
\begin{subfigure}{0.5\textwidth}
    \centering
    \includegraphics[width = 1\linewidth]{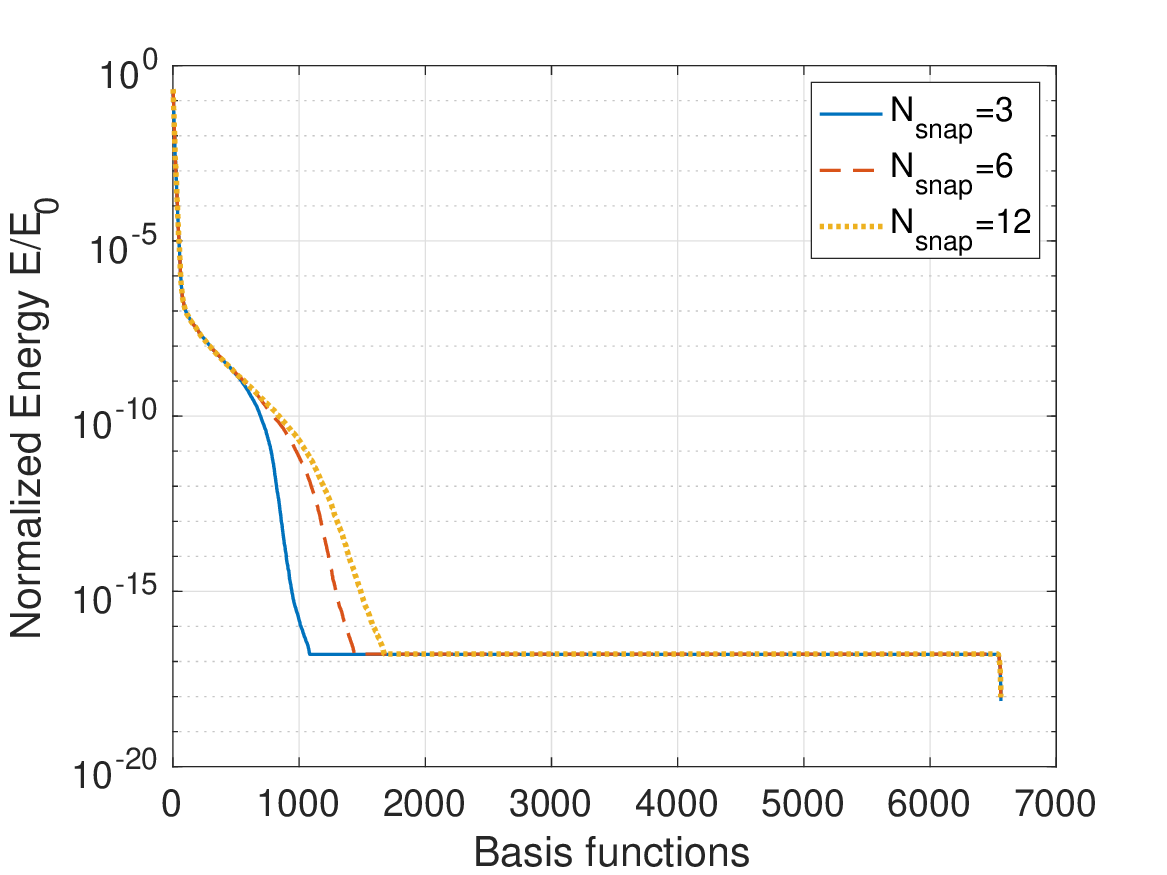}
    \caption{Energy decay for different number of sampled parameters. }
    \label{sfig:EnergtdecayAnaly6}
\end{subfigure} %

\caption{Illustration of computed singular value decay as a measure of the Kolmogorov n-width for different scenarios based on a square domain of size $2$ m $\times$ $2$ m with frequency-independent BC. $N_e=20$, $t=0.1$ s, $N_s = 3000$ and $N_{snap}=3$.}
\label{fig:EnergtdecayAnaly}

\end{figure}

To understand the RBM behavior in more details, we test additional conditions 1) different domain size in 2D, 2) number of complex frequencies in 2D, 3) number of sampled parameters. The effects of these changes on the singular value energy decay are studied in Figure \ref{fig:EnergtdecayAnaly}. Figure \ref{sfig:EnergtdecayAnaly1} shows the results when fixing the $PPW =10$ at $1000$ kHz. The number of basis functions needed to include a certain energy value decreases with increasing dimensions of the domain. In other words, the size of the ROM becomes smaller compared to the FOM when increasing the domain. Table \ref{Tab:percent} shows an example of the reduction in $DOF_{ROM}$ compared to $DOF_{FOM}$ for $\hat{\epsilon}_{POD}=10^{-6}$, which varies from from $45.2\%$ to $8.2\%$ when the edge length increases from $1$ m to $4$ m. The basis functions needed to capture the correct acoustics when making the domain larger do not increase as rapidly as the increase of the DOFs. 

In Figure \ref{sfig:EnergtdecayAnaly2}, the number of elements per direction is fixed to $N_e = 20$. When increasing the size of the domain, the number of basis functions needed increases due to the low resolution. Note that in such cases, the number of $PPW$ decreases when increasing the size of the domain. Figure \ref{sfig:EnergtdecayAnaly4} shows that the basis functions needed to include certain energy value increases when increasing $N_s$. The longer the simulation time is, the more complex frequencies are needed. Thus, more basis functions are needed to capture the wave propagation over longer time.

\begin{figure}[h]
\includegraphics[width=0.65\textwidth]{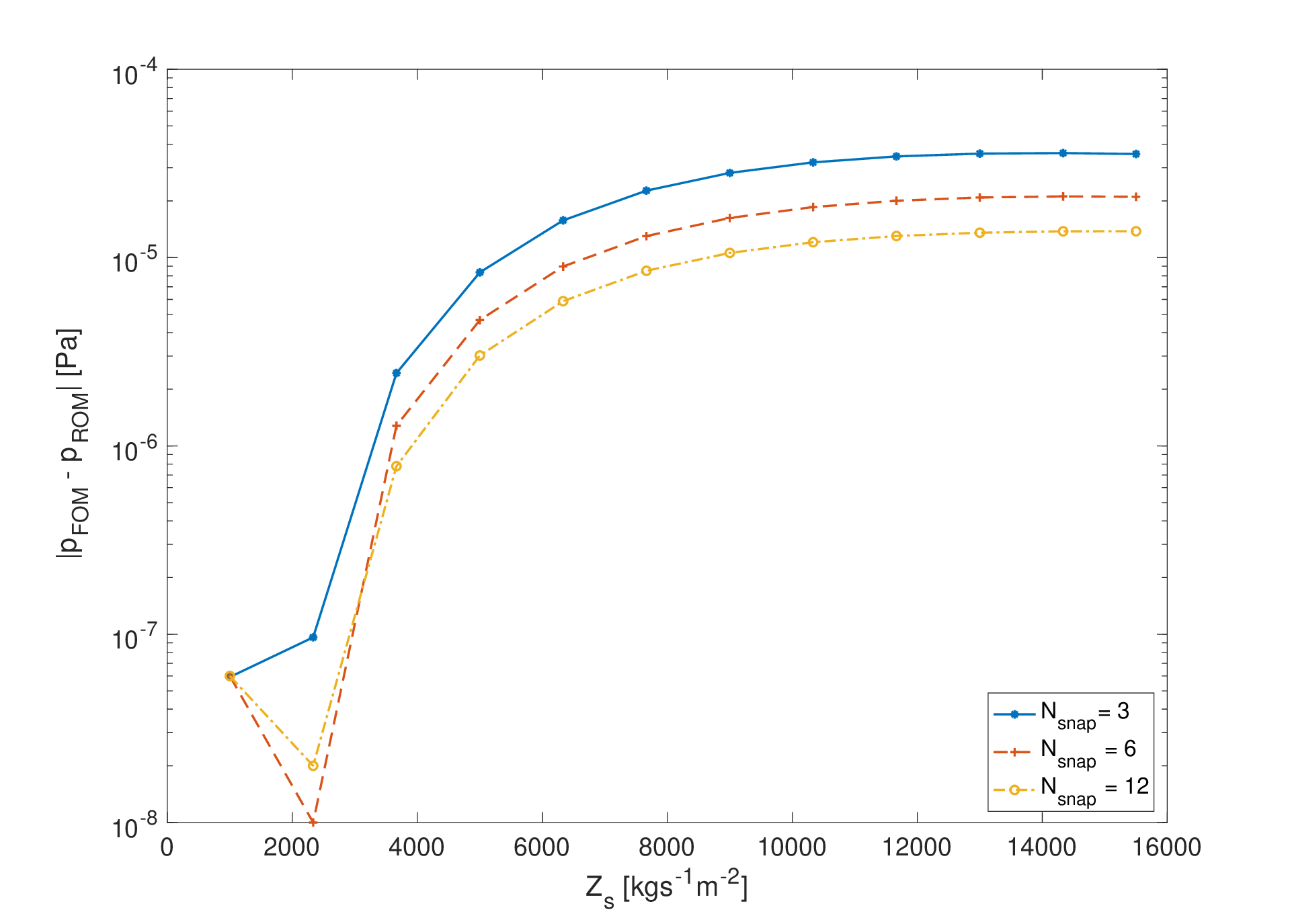}
\centering
\caption{\label{fig:sampleerror}{Error between ROM and FOM for different parameter values and $N_{snap}$ using $N_{rb}=44$ and $\hat{\epsilon}_{POD}=10^{-4}$.}}
\end{figure}

We investigated the effects of the number of offline calculations for the FOM parameter sampling, $N_{snap}$, on the energy decay in Figure \ref{sfig:EnergtdecayAnaly6}. A slower energy decay is observed when $N_{snap}$ is increased, meaning the the size of ROM gets larger. Because the upfront offline cost is expensive as $\mathcal{O}(N_{snap})$, we additionally investigated how the error of the ROM using the corresponding $N_{snap}$ varies as a function of $Z_s$ values in Figure \ref{fig:sampleerror} defined as $\epsilon=|p_{\text{FOM}}-p_{\text{ROM}}|$ computed at $t=0.1$ s in the receiver point ($r_x, r_y$)$=$($0.2, 0.2$) m. For most $Z_s$ values, the ROM with more umber of $N_{snap}$ are more accurate, but with a reasonable accuracy tolerance of $10^{-4}$, $N_{snap}$ of 3 is acceptable. For this particular case, increasing $N_{snap}$ decreases the error for most $Z_s$ values. However, it depends on the error tolerance. Moreover, the computational time of the offline stage increases with $\mathcal{O}(N_{snap})$. The results show that for a uniform sampling strategy, increasing $N_{snap}$ produces a slower energy decay of the singular values (Figure \ref{sfig:EnergtdecayAnaly6}), an increased accuracy (Figure \ref{fig:sampleerror}) and an increase of the offline computational time. It can be seen that with $N_{snap}=3$ the acoustics of the system can be captured accurately to cover the whole practical range of interest of the parametrized surface impedance. This suggests that the RBM approach is more beneficial when solving large scale problems, which is the case in room acoustics.

\begin{table}[h!]
\caption{$DOF_{ROM}/DOF_{FOM}$ in percentage for $\hat{\epsilon}_{POD}=1\times10^{-6}$.} 
\label{Tab:percent}
\centering
\small
\begin{tabular}{lllll}
\\ \hline \hline
\multicolumn{1}{l|}{Domain side length}  & 1 m & 2 m & 3 m& 4 m \\ \hline
\multicolumn{1}{l|}{$DOF_{ROM}/DOF_{FOM}$} & 45.2\% & 12.3\% & 11.9\% & 8.2\% \\ \hline \hline
\end{tabular}
\end{table}

The same analysis has been done for frequency-dependent boundaries. Results are similar to those as described for the frequency-independent case.

\subsubsection{Computational cost and storage}

\begin{figure}[h]

\begin{subfigure}{0.5\textwidth}
    \centering
    \includegraphics[width = 1\linewidth]{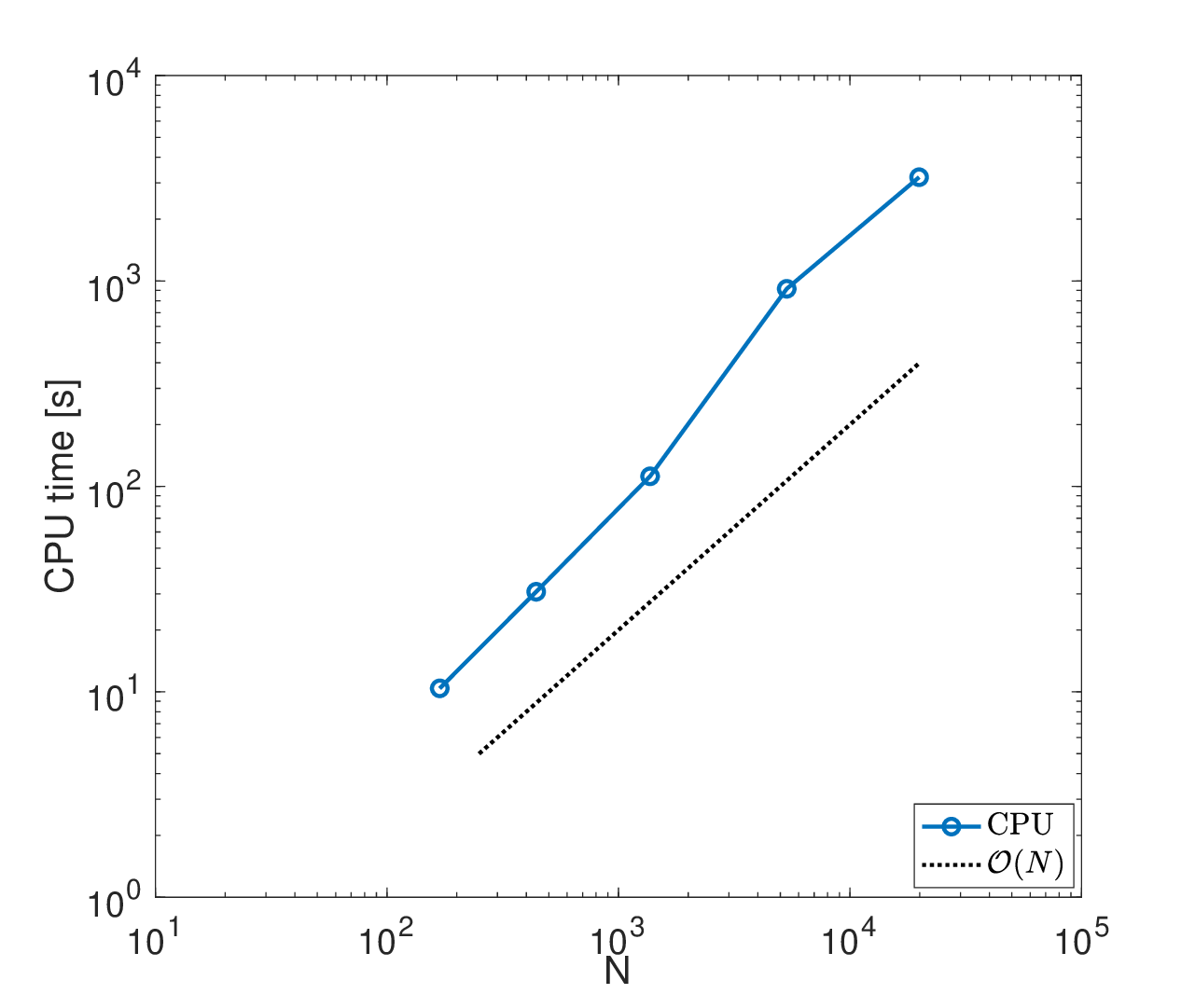}
    \caption{CPU time of the FOM for different degree of freedom.}
    \label{fig:fdepcost1}
\end{subfigure} %
\begin{subfigure}{0.5\textwidth}
    \centering
    \includegraphics[width = 1\linewidth]{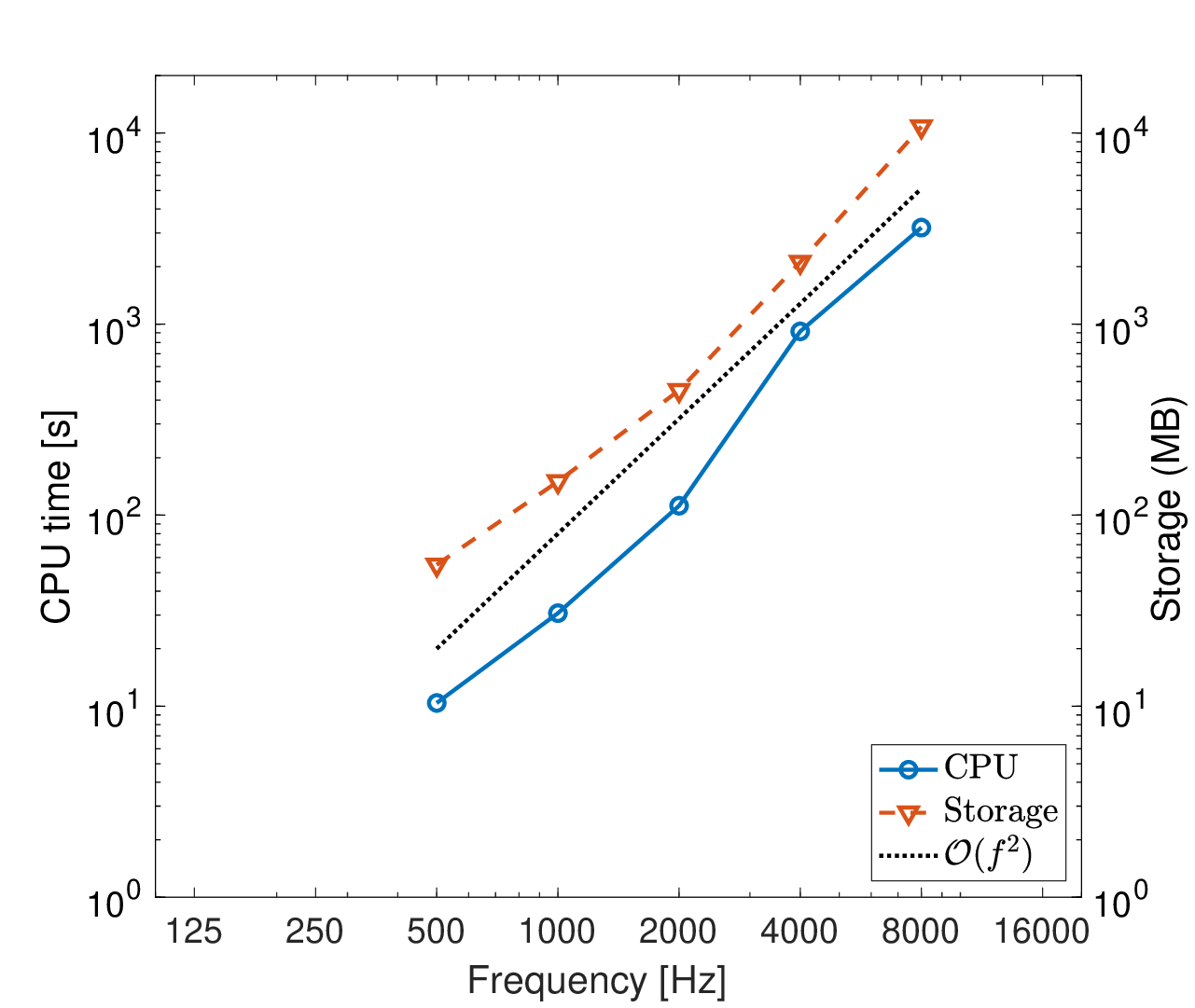}
    \caption{CPU and storage of the ROM for different frequency limits.}
    \label{fig:fdepcost2}
\end{subfigure}
\caption{\label{fig:fdepcost}{Computational and storage cost for frequency-dependent 2D case. Simulations were carried out with a fixed $PPW=6$ and $P=4$.}}
\end{figure}

An analysis of the computational and storage cost of the previously presented 2D ROM with frequency-dependent boundary conditions is performed.  Figure \ref{fig:fdepcost1} presents the computational time of the FOM for different DOF where it can be seen that the CPU time increases when increasing $N$ as $\mathcal{O}(N)$. Moreover, Figure \ref{fig:fdepcost2} shows the computational cost in seconds for each FOM sampled parameter as a function of the upper limit frequency given by the element size. For this analysis, a polynomial order of $P=4$ and a six-elements per wavelength thumb rule is considered \cite{sixelem}. Note that in order to know the total cost of all the snapshots, the values need to be multiplied by the number of snapshots. The results show that the CPU time is increasing as  $\mathcal{O}(f^2)$. Moreover, the storage of ROM including all the snapshots is also plotted, which increases as a quadratic function of $f$. The dependency with the number of complex frequencies $N_s$ is straightforward. Changing $N_s$ by a factor $x$ will scale results in Figure \ref{fig:fdepcost} $x$ times. 

\section{Analysis and discussion}


The ROM developed for room acoustic simulations concurs with previous work from other disciplines, e.g., \cite{scatteringRBM,rozzaappliRBM,PhuongDavidRBM}, in terms of speedup factors with respect to the error. A speedup of 100 is found with an error level of $6\times 10^{-8}$ Pa and $2\times 10^{-5}$ Pa for 2D frequency-independent and frequency-dependent, respectively. For the 3D frequency-dependent case, a speedup of 300 is found when the error is $6\times 10^{-4}$ Pa and 1300 for an error value equal to $9.8\times 10^{-3}$ Pa. It is clear that a compromise between the desired error and the speedup occurs. In room acoustics, the impulse response is used to characterize a room for a given source and receiver position. One of the more important acoustic parameters is the reverberation time, $T_{30}$ or $T_{20}$, defined in ISO 3382-1 \cite{ISO3382-1}, which can be obtained from the impulse response and is commonly used for evaluating the acoustic conditions of an enclosure. The precision needed to calculate reverberation times is related to the dynamic range of the signal given by the background noise. For $T_{30}$ and $T_{20}$, a background noise level of $-45$ dB and $-35$ dB respectively is needed \cite{ISO3382-1}, which correspond to errors of $5.6\times 10^{-3}$ and $1.77\times 10^{-2}$. For sound pressure level (SPL) prediction, the just noticeable difference for a broadband noise is 1 dB, which amount to about 11\% error.




Results show a potential speedup when increasing the size of the problem or $N$ as shown in Figure \ref{fig:ROMSPEEDUP}, so mostly beneficial for large 3D domains. Figure \ref{fig:fdepcost1} shows how the computational time of the 2D FOM increases when $N$ grows. On the other hand, increasing $N$ could be a problem in terms of memory storage. However, Figure \ref{fig:fdepcost2} shows affordable storage values for the presented cases. It is found that for the analyzed 2D case, 2GB of storage is needed for an upper frequency of 4 kHz and 10.8GB for 8 kHz. Moreover, the singular value decay analysis shows a potential benefit for large rooms Figure \ref{fig:EnergtdecayAnaly}. In the analogous 3D case, 8GB of storage is needed for an upper frequency of 1kHz. All these results show favorable conditions for room acoustics with reasonable domain sizes.

 
 The main drawback of the current work is that constructing the impulse response is time-consuming due to the parameter search for the Weeks method. The optimal choice of the Weeks parameter also depends on $N_s$, and it is case dependent. It is relieving that larger optimal parameter ranges exist for non-rigid boundary condition, as many practical building materials are acoustically non-rigid. 
 
 A future work is to investigate this further and develop a more efficient method to estimate the parameters. Moreover, the framework will be extended in future work by focusing on larger domains for realistic 3D cases. Parallel and high-performance computing techniques are also candidates to overcome the extra computational cost for those complex and larger scenarios. 


\section{Conclusion}
 This study evaluates the potential of using RBM techniques for parametrized boundaries in room acoustic simulations. The proposed framework reduces the problem to a low dimensional subspace aiming a computational reduction using a Laplace domain ROM on a high fidelity SEM solver, including complex boundary conditions. The solution is finally transformed back to the time-domain to reconstruct the impulse responses. The test cases presented in this paper show the potential of this framework for room acoustics simulations. A typical scenario is to simulate iteratively a room for different absorption properties of the boundaries to find the optimal values that fulfil a pre-defined acoustic requirement. It is shown that the ROM enables simulating different parameter values of the boundaries at a significantly lower cost compared to the FOM. Results confirm that the use of RBM decreases the simulation time by at least two order of magnitude for the 2D cases presented here and three order of magnitude for the 3D cases, including the frequency-independent and dependent boundary conditions. The speedup including the offline stage, where the ROM is constructed by sampling the parameter variations on the FOM, is three orders of magnitude faster than the FOM when four different boundary conditions are simulated per room surface in a 3D case with frequency-dependent boundaries. 
 
Results show favourable behaviour in terms of speedup when increasing the size of the domain, indicating a potential further accelerations for larger domains. Moreover, a memory storage is manageable for the tested cases, albeit small, for a unit cube. This suggests that the presented framework is expected to be most useful for larger scale engineering problems, such as building design and renovation.

A challenge is on the choice of the number of complex frequencies and the free parameters to construct the impulse response by means of Weeks method. However, results show that for a room with damped boundaries, broader ranges of the Weeks parameters ensure good accuracy, which is relevant for room acoustics as absorption at the boundaries is always included.

\section*{Acknowledgments}
This work is partly supported by Innovationsfonden, Denmark
(Grant ID 9065-00115B), Rambøll Danmark A/S and Saint-Gobain Ecophon A/S, Sweden.

\bibliography{Manuscript}

\newpage
\mbox{}
\nomenclature{\(c\)}{Speed of the sound}
\nomenclature{\(\hat{p}\)}{Sound pressure in time-domain}
\nomenclature{\(t\)}{Time}
\nomenclature{\(T\)}{Final time}
\nomenclature{\(\Omega\)}{Domain}
\nomenclature{\(\boldsymbol{x}\)}{Cartesian coordinates in the domain}
\nomenclature{\(\boldsymbol{\hat{v}}\)}{Particle velocity in time-domain}
\nomenclature{\(\rho\)}{Density of the medium (air)}
\nomenclature{\(\sigma_g\)}{Spatial variance of the Gaussian pulse}
\nomenclature{\(s\)}{Complex frequency of the Laplace domain}
\nomenclature{\(\sigma\)}{Real part of the complex frequency}
\nomenclature{\(\gamma\)}{Imaginary part of the complex frequency}
\nomenclature{\(p\)}{Sound pressure in the Laplace domain}
\nomenclature{\(\hat{p}_0\)}{Initial condition in time-domain}
\nomenclature{\(\hat{p}_{t_0}\)}{Sound pressure derivative in t=0 s}
\nomenclature{\(\boldsymbol{x}_0\)}{Source position for the initial condition}
\nomenclature{\(\boldsymbol{v}\)}{Particle velocity in Laplace domain}
\nomenclature{\(w\)}{Test function}
\nomenclature{\(\Gamma\)}{Boundary of the domain $\Omega$}
\nomenclature{\(\boldsymbol{n}\)}{Outward pointing normal vector of the boundary}
\nomenclature{\(Z_s\)}{Surface impedance}
\nomenclature{\(p_{\Gamma}\)}{Sound pressure in the Laplace domain at the boundary}
\nomenclature{\(v_{n}\)}{Normal velocity at the boundary}
\nomenclature{\(\mathcal{M}\)}{Mass matrix}
\nomenclature{\(\mathcal{M}_{\Gamma}\)}{Mass matrix for the boundary}
\nomenclature{\(\mathcal{S}\)}{Stiffness matrix}
\nomenclature{\(\mathcal{S}_x, \mathcal{S}_y \mathcal{S}_z\)}{Stiffness matrix, where x,y,z denote differentiation}
\nomenclature{\(\boldsymbol{p}\)}{Sound pressure vector in Laplace domain}
\nomenclature{\(\boldsymbol{K}\)}{Matrix operator} 
\nomenclature{\(\mathbf{G}\)}{Matrix operator for the accumulators}
\nomenclature{\(\boldsymbol{Q}\)}{Right hand side vector}
\nomenclature{\(\mathcal{B}\)}{Boundary matrix}
\nomenclature{\(Y_s\)}{Surface admittance}
\nomenclature{\(a_Q\)}{Zeros of the rational function}
\nomenclature{\(b_Q\)}{Poles of the rational function} 
\nomenclature{\(Q\)}{Order of the rational function} 
\nomenclature{\(L\)}{Number of real poles} 
\nomenclature{\(\hat{\lambda}\)}{Real poles} 
\nomenclature{\(S\)}{Number of complex poles} 
\nomenclature{\(\hat{\alpha}_k,\hat{\beta}_k\)}{Complex poles} 
\nomenclature{\(A_k,B_k,C_k\)}{Coefficients} 
\nomenclature{\(\hat{\phi}_k,\hat{\psi}_k^{(1)},\hat{\psi}_k^{(2)}\)}{Accumulators} 
\nomenclature{\(\boldsymbol{G}\)}{Matrix operator for accumulators}
\nomenclature{\(L_k(\cdot)\)}{Laguerre polynomials of degree $k$}
\nomenclature{\(\hat{a}_k\)}{Coefficients for Weeks method}
\nomenclature{\(b\)}{Week free parameter}
\nomenclature{\(N\)}{Number of degree of freedom}
\nomenclature{\(N_s\)}{Number of complex frequencies}
\nomenclature{\(s_j\)}{$j$th complex frequency}
\nomenclature{\(b^{opt}\)}{Optimal value of the free parameter}
\nomenclature{\(\sigma^{opt}\)}{Optimal value of the free parameter}
\nomenclature{\(\hat{p}^*_i\)}{Solution simulated with time-domain solver}
\nomenclature{\(p^*_i\)}{Solution simulated with Laplace domain solver}
\nomenclature{\(N_{rb}\)}{Number of reduced basis}
\nomenclature{\(N_{s}\)}{Number of complex frequencies $s$}
\nomenclature{\(p_{FOM}\)}{High-fidelity solutions}
\nomenclature{\(\textit{M}\)}{Solution Manifold}
\nomenclature{\(\mu\)}{Parameter for RBM}
\nomenclature{\(p_{ROM}\)}{Reduced basis solution}
\nomenclature{\(\phi_i\)}{Reduced basis}
\nomenclature{\(a_i\)}{Reduced basis solution coefficient}
\nomenclature{\(\boldsymbol{S_N}\)}{Snapshot matrix}
\nomenclature{\(\boldsymbol{S}_{\boldsymbol{N}cl}\)}{Simplectic snapshot matrix}
\nomenclature{\(\boldsymbol{\Psi}\)}{Truncated reduced basis}
\nomenclature{\(\boldsymbol{\Psi}\)}{Simplectic truncated reduced basis}
\nomenclature{\(E/E_0\)}{Normalized energy of the singular values}
\nomenclature{\(\delta\)}{Singular values}
\nomenclature{\(I(N_{rb})\)}{Indicator for determining $N_{rb}$ for a given tolerance $\hat{epsilon}_{POD}$}
\nomenclature{\(\boldsymbol{K}_{rb}\)}{Reduced matrix operator}
\nomenclature{\(\boldsymbol{q}_{rb}\)}{Reduced right hand side vector}
\nomenclature{\(\mathcal{M}_{\Phi}\)}{Reduced mass matrix}
\nomenclature{\(\mathcal{B}_{\Phi}\)}{Reduced boundary matrix}
\nomenclature{\(\mathcal{S}_{\Phi}\)}{Reduced stiffness matrix}
\nomenclature{\(\alpha_{norm}\)}{Normal incidence absorption coefficient}
\nomenclature{\(P\)}{Polynomial order for SEM}
\nomenclature{\(\Delta t\)}{Time step}
\nomenclature{\(N_e\)}{Number of elements per direction}
\nomenclature{\((r_x,r_y,r_z)\)}{Receiver position}
\nomenclature{\((s_x,s_y,s_z)\)}{Source position}
\nomenclature{\(\sigma_{mat}\)}{Flow resistivity of porous material}
\nomenclature{\(d_{mat}\)}{Porous material thickness}
\nomenclature{\(N_t\)}{Number of time steps}
\nomenclature{\(N_{snap}\)}{Number of sampled parameters}

\printnomenclature

\end{document}